\documentclass[floatfix]{iopart}
\usepackage{iopams} 
\usepackage{setstack}
\usepackage{geometry}
\usepackage{color}
\usepackage{amssymb}
\usepackage{mathrsfs}

\pdfoutput=1

\usepackage{graphicx,epsfig}
\usepackage{bm}

\usepackage{braket}

\usepackage{color}
\usepackage{comment}
\definecolor{darkred}{rgb}{0.8,0.1,0.1}
\def\normOrd#1{\mathop{:}\nolimits\!#1\!\mathop{:}\nolimits}
\begin{document}

\paper[R\'enyi entanglement entropies of descendant states in critical systems with boundaries]{R\'enyi entanglement entropies of descendant states in critical systems with boundaries: conformal field theory and spin chains}

\author{Luca Taddia}
\address{Scuola Normale Superiore, Piazza dei Cavalieri 7, I-56126 Pisa, Italy}
\address{CNR - Istituto Nazionale di Ottica, Sede Secondaria di Sesto Fiorentino, Via Carrara 1, I-50019 Sesto Fiorentino, Italy}
\ead{luca.taddia2@gmail.com}

\author{Fabio Ortolani}
\address{Dipartimento di Fisica e Astronomia  dell'Universit\'a  di  Bologna \& INFN, Sezione di Bologna,  Via  Irnerio  46,  I-40127  Bologna,  Italy}

\author{Tam\'as P\'almai}
\address{MTA-BME ``Lendulet'' Statistical Field Theory Research Group, Institute of Physics, Budapest University of Technology and Economics, Budafoki ut 8, H-1111 Budapest, Hungary}

\begin{abstract}
We discuss the R\'enyi entanglement entropies of descendant states in critical one-dimensional systems with boundaries, that map to boundary conformal field theories in the scaling limit. We unify the previous conformal-field-theory approaches to describe primary and descendant states in systems with both open and closed boundaries. We provide universal expressions for the first two descendants in the identity family. We apply our technique to critical systems belonging to different universality classes with non-trivial boundary conditions that preserve conformal invariance, and find excellent agreement with numerical results obtained for finite spin chains. We also demonstrate that entanglement entropies are a powerful tool to resolve degeneracy of higher excited states in critical lattice models.
\end{abstract}

\section{Introduction}

Understanding quantum correlations and entanglement in many-body systems is one of the main purposes of fundamental physics \cite{Amico2008}. Although a general strategy for this task currently lacks, in the last decades many advances have been made. In particular, one-dimensional (1D) systems play a very special role in this scenario, for two reasons: the first is of physical nature, and resides in the enhancement of the importance of quantum fluctuations, due to dimensionality \cite{Giamarchi2003}; the second is the existence of extremely powerful analytical and numerical techniques, such as exact solutions \cite{Sachdev2011}, bosonization \cite{Giamarchi2003}, Bethe ansatz \cite{Takahashi1999}, and matrix-product-states algorithms \cite{Schollwock2011}, allowing for the extraction of accurate information about low-lying excitations in a non-perturbative way. 
Within the bosonization framework, the strategy is to identify the relevant degrees of freedom of the considered 1D model and, starting from them, to build up an effective field theory capturing its low-energy physics \cite{Giamarchi2003}. If the ground state (GS) of the original model is gapless, the obtained field theory is usually conformally invariant, and is therefore called conformal field theory (CFT) \cite{Henkel1999,DiFrancesco1997} (we remark that bosonization is not the only approach leading to effective CFT's; see, e.g., Ref. \cite{Mussardo2010}). Due to their solvability in $(1+1)$D, CFT's are particularly useful, allowing for the exact computation of all correlation functions, and therefore providing access to many interesting quantities in a controllable way \cite{Belavin1984}.

In the present work we deal with a particular class of entanglement measures, the R\'enyi entanglement entropies (REE's) \cite{Renyi2007,Nielsen2000}, defined in the following way. Let us consider a pure state of an extended quantum system, associated with a density matrix $\hat\rho$, and a spatial bipartition of the system itself, say $\left\{A,B\right\}$. If we are interested in computing quantities that are spatially restricted to $A$, we can employ, instead of the full density matrix, the reduced density matrix $\hat\rho_A=\mbox{Tr}_B\hat\rho$. The $n$-th REE, defined as
\begin{equation}
S_n(A)=\frac{1}{1-n}\log_2\mbox{Tr}_A\hat\rho_A^n,
\end{equation}
describes the reduced density matrix. In the limit $n\rightarrow1$ it reproduces the von Neumann entanglement entropy (VNEE) $S(A)=-\mbox{Tr}_A\left[\hat\rho_A\log_2\hat\rho_A\right]$, the most common entanglement measure \cite{Nielsen2000}. In the last decade, the $n\neq1$ REE's have also become quite popular, for several reasons. Analytical methods proved to be more suitable to calculate REE's than VNEE, especially in field theory: e.g., in CFT, REE's have a clear interpretation as partition functions, while the VNEE does not \cite{Holzhey1994,Calabrese2004}. From a fundamental point of view, the knowledge of the REE's $\forall n\in\mathbb{N}$ is equivalent to the knowledge of $\hat\rho_A$ itself, since they are proportional to the momenta of $\hat\rho_A$. At the same time, from a physical point of view, many of the important properties of the VNEE, e.g., the area law for gapped states \cite{Eisert2010,Hastings2007,Huang2014} and the proportionality to the central charge for critical systems \cite{Vidal2003,Holzhey1994,Calabrese2004}, carry over to REE's as well. In fact, REE's are easily computable by matrix-product-states algorithms \cite{Schollwock2011}, and they allow for a precise estimation of the central charge from numerical data regarding the GS of the system (see, e.g., Ref. \cite{Barbarino2015}). Finally, and very importantly, measurements of the $n=2$ REE in an ultracold-atoms setup have been recently performed \cite{Islam2015,Kaufman2016}, paving the way to the experimental study of entanglement measures in many-body systems.

The computations of Refs. \cite{Holzhey1994,Calabrese2004}, that deal with the GS of CFT's, have been extended, in more recent times, to excited states \cite{Alcaraz2011,Berganza2012,Taddia2013,Palmai2014}. In CFT, excitations are in one-to-one correspondence with fields, and can be organized in conformal towers \cite{DiFrancesco1997}. The lowest-energy state of each tower is in correspondence with a so-called {\it primary} field, while the remaining ones, called {\it descendant}, are in correspondence with {\it secondary} fields, obtained from the primaries by the application of conformal generators. While the properties of REE's for primary states are now quite well understood \cite{Alcaraz2011,Berganza2012,Taddia2013}, much less is known about the descendants. In the pioneering work of Ref. \cite{Palmai2014}, a unifying picture for the computation of REE's of primary and descendant states was developed and both analytical and numerical computations for the scaling limit of simple spin chains with periodic boundary conditions (PBC) were performed. A related but physically different problem was considered in Refs.  \cite{Caputa2015,Chen2015}, where the effect on the time evolution of inserting secondary operators at finite time was studied.

In this paper we continue the work in Ref. \cite{Palmai2014} and extend its framework to the case of open boundary conditions (OBC) that preserve conformal invariance. This is an important step, for several reasons. First, impurity problems \cite{Affleck2000} and certain problems in string theory \cite{Polchinski1998} map to boundary CFT; not secondarily, the experimentally achievable setups often involve OBC.
The importance of descendant states stems from novel applications, including non-trivial checks of the universality class of critical lattice models, and understanding the behavior of degenerate multiplets.
We will provide a general strategy for the computation of REE's, derive CFT predictions for descendant states, and compare them to the numerical data obtained from lattice realizations of the considered CFT's. We will also discuss some interesting complementary aspects arising naturally when considering descendant states: e.g., we will need to consider the REE's of linear combinations on the CFT side, in order to study degenerate states in the XX chain.

The paper is structured as follows. In Section \ref{CFT_sec} we will review the approach of Ref. \cite{Palmai2014} for systems with PBC, derive the procedure to compute the REE's for descendant states in CFT's with (and without) boundaries, and provide some general results for states related to the tower of the identity. In Section \ref{Ising_sec} we will compute explicitly the REE's for the $c=1/2$ minimal CFT, and compare the scalings to the numerical data obtained for the spin-1/2 Ising model in a transverse field; in Section \ref{Potts} we will perform the same study for the $c=4/5$ minimal CFT and the three-state Potts model, while in Section \ref{XX} for the compactified free boson and the spin-1/2 XX chain. In Section \ref{conc} we will draw our conclusions and some directions for future work. In the Appendixes we will provide useful technical details of our discussion.

\section{Results from CFT}\label{CFT_sec}

In the scaling limit, critical 1D lattice models can be described by CFT's \cite{Giamarchi2003,Mussardo2010}. Since CFT's are exactly solvable, explicit universal expressions can be derived for the REE's. In this Section we describe a calculation scheme for the REE's of arbitrary excited states in CFT unifying the cases of PBC and OBC. We consider here only unitary theories; see Ref.  \cite{Bianchini2015} for a generalization to the ground states of non-unitary models.

\subsection{Periodic boundary conditions}

The case of a finite system with PBC has been widely studied in the past 
(see, e.g., Refs. \cite{Holzhey1994, Calabrese2004, Alcaraz2011, Berganza2012, Palmai2014}). In this Section, we review the computation of the REE's for excited 
states in this case, following the approach of Refs. \cite{Berganza2012, Palmai2014} and modifying it slightly, having in mind the generalization to the OBC case.

We consider an Euclidean space-time manifold of an infinite cylinder: it is characterized by the complex variable $r-i\tau$, being $r$ a space coordinate, and $\tau$ a time one; $r-i\tau$ and $r+L-i\tau$, where $L$ is the size of the system (acting as an IR cutoff for the field theory), are identified. The physical support is the circle at $\tau=0$; the subsystem $A$ is chosen to be the interval $[-\ell/2,\ell/2]$. The zero-temperature density matrix of the system is pure, i.e., $\hat\rho_{\Psi}=\left|\Psi\right>\left<\Psi\right|$, where $\left|\Psi\right>$ is a generic eigenstate of the Hamiltonian.

The starting point of our computation is the following identity:
\begin{equation}
\fl\Tr_A\rho^n_{A,\Psi}\label{replica_eq}
=\sum_{a_1b_1}\sum_{a_2b_2}\ldots\sum_{a_nb_n} 
\langle a_1b_1\vert\Psi\rangle\langle\Psi\vert a_2 b_1\rangle 
\langle a_2b_2\vert\Psi\rangle\langle\Psi\vert a_3 b_2\rangle\ldots 
\langle a_nb_n\vert\Psi\rangle\langle\Psi\vert a_1 b_n\rangle,
\end{equation}
where we repeatedly inserted resolutions of the identity for the Hilbert spaces 
relative to the segments $A$ and 
$B$. Because of the 
state-operator correspondence in CFT, each eigenstate is generated by an 
operator $\Psi$, acting on the vacuum and placed at the infinite past 
\cite{DiFrancesco1997}. Adopting this representation, the overlaps above are nothing 
but path integrals on half of the infinite complex cylinder, with the insertions of $\Psi$ and $\Psi^\dagger$ in the far past and future respectively, and with boundary 
states $a_i$ and $b_j$ along the segments $A$ and $B$. 
When the sums in Eq. (\ref{replica_eq}) are performed we obtain a 
correlation function of $\Psi$ and $\Psi^\dagger$ operators on the so-called {\it replica 
manifold}, consisting of $n$ copies of the cylinder, glued together cyclically
across cuts along $A$ \cite{Holzhey1994,Calabrese2004}. Each of the $n$ copies 
can be transformed to the complex plane by the conformal mapping 
$\xi_j=e^{i\frac{2\pi}{L}(r_j-i\tau_j)}$ \cite{DiFrancesco1997}: the resulting 
$n$-sheeted manifold, that will be denoted by $\mathcal{R}_n$, is a 
collection of $n$ planes glued across the boundaries 
$\left(\mathcal{A}\pm a\right)_j$ according to 
$\left(\mathcal{A}-a\right)_j\leftrightarrow\left(\mathcal{A}+a\right)_{j+1}$, where
$\mathcal{A}$ is the arc of the unit 
circle $e^{i\phi}$, $\phi\in[-\pi x,\pi x]$ ($x=\ell/L$ is the relative subsystem size), and $a$ is a radial infinitesimal vector (related to, e.g., the lattice spacing in the case of a lattice model), introduced in order to 
UV-regularize the theory.

\begin{figure}[t]
\centering
\includegraphics[width=0.45\linewidth]{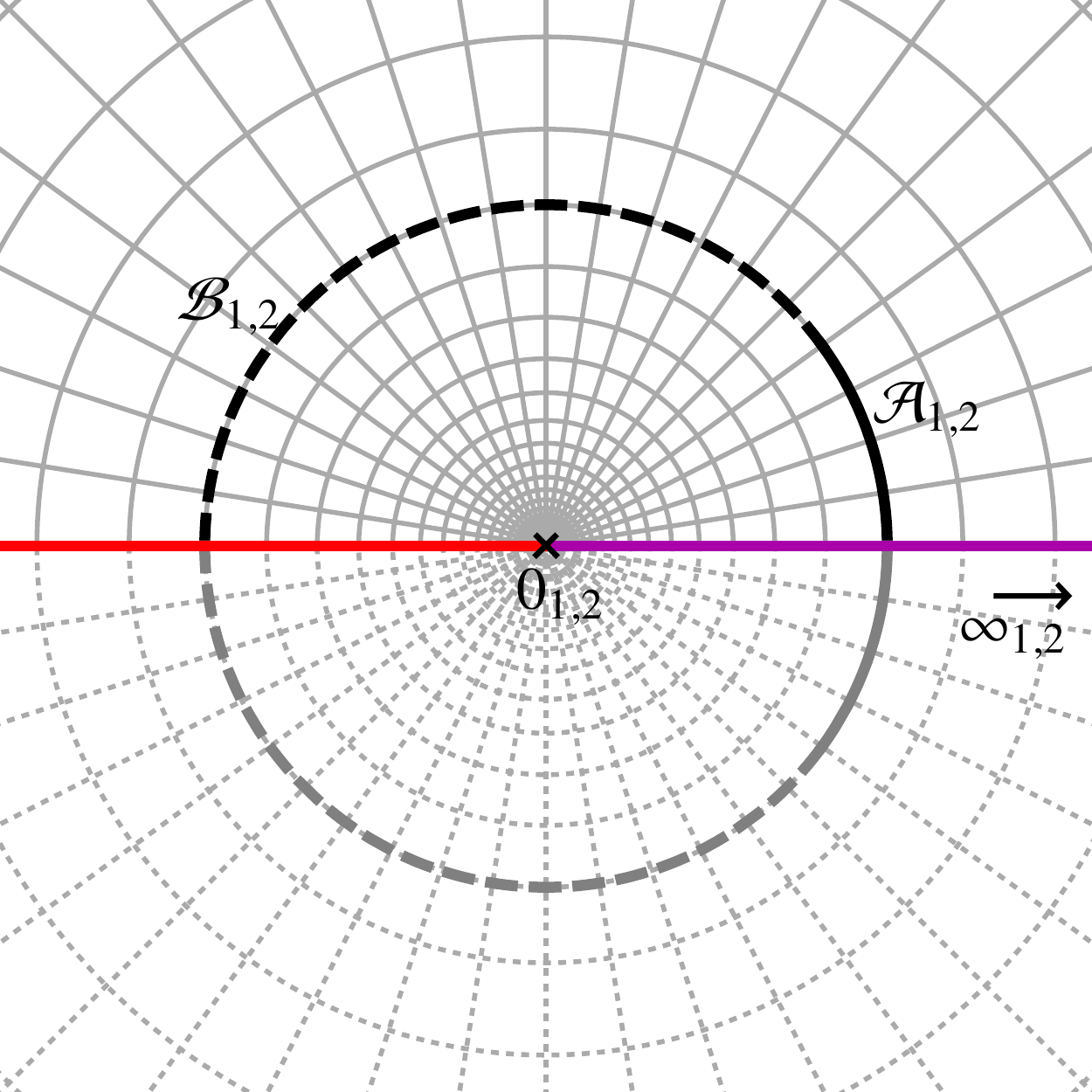}
$\quad$\includegraphics[width=0.45\linewidth]{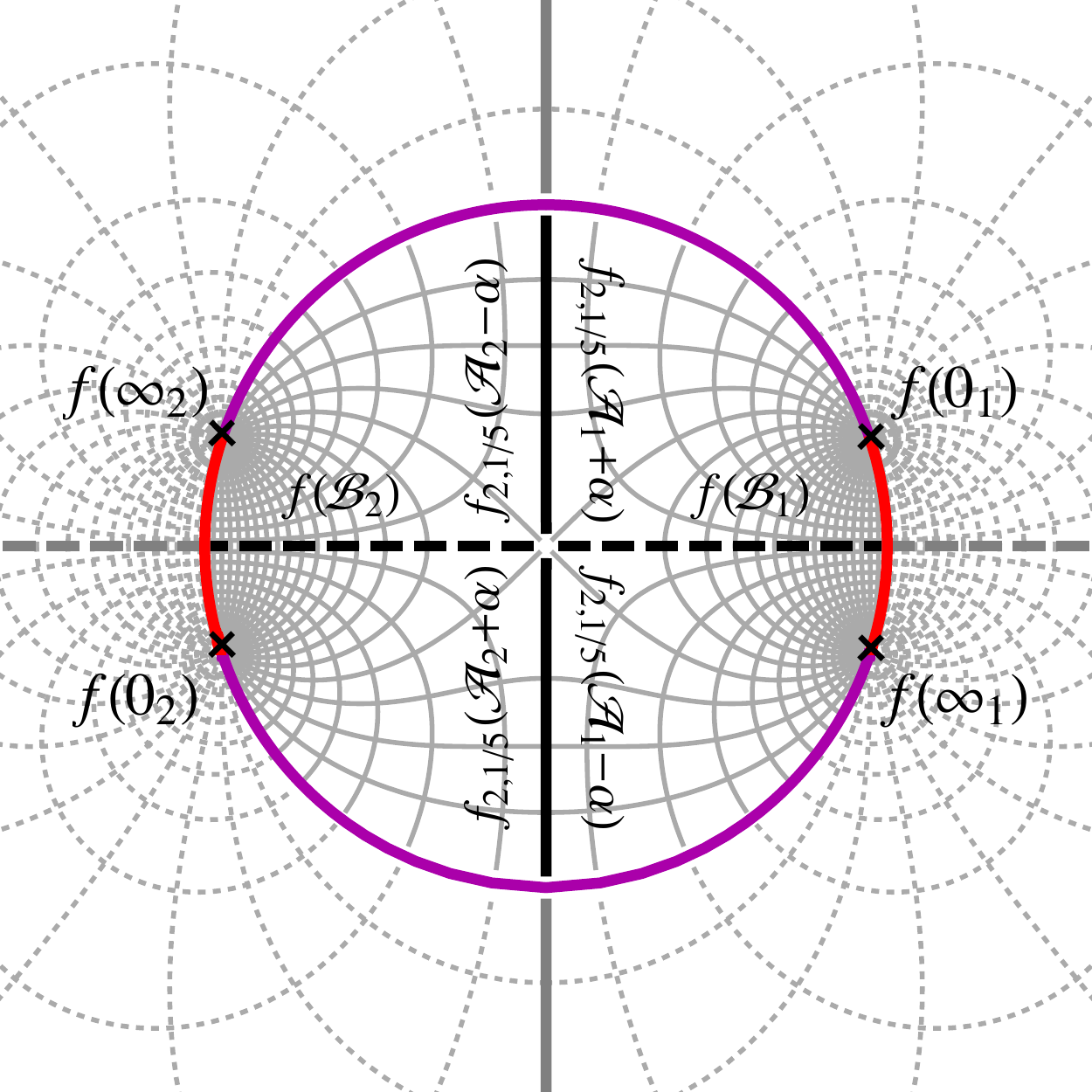}
\caption{Conformal map $f_{2,1/5}$ regularizing the replica manifold for both PBC and OBC. The mesh represent lines of equal time and space coordinates. In the OBC case only the solid lines are in the physical domain. The purple (red) line represents the left (right) edge of the OBC strip. Left panel: pre-transformed space-time. Right panel: transformed space-time.}\label{transf}
\end{figure}

Assuming the usual normalization for the state, i.e., 
$\langle\Psi\vert\Psi\rangle_{\mathcal{R}_1}=1$, $\Tr_A\rho_{A,\Psi}^n$ 
reduces to a $2n$-point function on the replica manifold, that we then 
transform into a $2n$-point function on a single plane (the following relation contains a non-trivial statement: see the end of this Section for a clarification of this point): 
\begin{equation}\label{path-int}
\fl\Tr_A\rho_{A,\Psi}^{n}  = 
\mathcal{N}_{n}\left<\prod_{j=1}^{n}\Psi\left(0_{j},0_j\right)
\Psi\left(0_{j},0_j\right)^{\dagger}\right>_{\mathcal{R}_{n}} 
=\mathcal{N}_{n}\left<\prod_{j=1}^{n}\mathcal{T}_{f_{n,x}}\Psi\left(0_{j},0_j\right)
\mathcal{T}_{f_{n,x}}\Psi\left(0_{j},0_j\right)^{\dagger}
\right>_{\mathbb{C}},
\end{equation}
where the constant $\mathcal{N}_n=Z_n/Z_1^n$ ($Z_n$ is the partition function 
over $\mathcal{R}_n$) sets the correct normalization $\Tr_A\rho_{A,\Psi}=1$, and 
takes, in the present case, the form
\begin{equation}
\mathcal{N}_{n}=\left[\frac{L}{\pi a}\sin\left(\pi 
x\right)\right]^{\frac{c}{6}\left(\frac{1}{n}-n\right)},
\end{equation}
where $c$ is the central charge of 
the considered CFT \cite{DiFrancesco1997}. The second expression in Eq. 
(\ref{path-int}) is obtained by means of the conformal mapping $\xi\to 
f_{n,x}(\xi)$ from $\mathcal{R}_n$ to the complex plane, i.e., a composition of 
a M\"obius transformation and the $n$-th root:
\begin{equation}\label{CT}
f_{n,x}(\xi)=\left(\frac{e^{-i\pi x}\xi-1}{\xi-e^{-i\pi x}}\right)^{1/n}.
\end{equation}
The M\"obius transformation brings the cut along the arc $\mathcal{A}$ to the 
half line $(-\infty,0]$; then, the $n$-th root transforms each replica sheet to 
a slice of the complex plane. We prescribe the $j$-th sheet 
($j=1,\ldots,n$) to be transformed by the $j$-th branch of the 
$n$-th root.
The mapping for $n=2$ is represented graphically in Figure \ref{transf}. Finally, 
the operator $\Psi(z)$ is transformed into 
$\mathcal{T}_{f_{n,x}}\Psi(z)$ under the mapping (\ref{CT}). The symbol $\mathcal{T}$ is not to be confused with the twist operator appearing when the R\'enyi entropy is expressed as a quantity in a local theory \cite{Cardy2007}. The transformed operator is inserted at the point $f_{n,x}(z)$; for details see \ref{appA}.

The normalization constant can be identified (apart from a non-universal additive constant; see Eq. (\ref{CCREE})) with the GS contribution
\begin{equation}\label{logNn}
 S_{n}^{\rm{GS}}=\frac{1}{1-n}\log_2\mathcal{N}_n;
\end{equation}
then if we rewrite the REE as
\begin{equation}
S_{n}^{\Psi}=S_{n}^{\rm{GS}}+\frac{1}{1-n}\log_2F^{(n)}_{\Psi},
\end{equation}
the $n$-th (exponentiated) {\it excess entanglement entropy} (EEE) $F^{(n)}_{\Psi}$ corresponds to the $2n$-point function
\begin{equation}
F^{(n)}_\Psi=\left<\prod_{j=1}^{n}\mathcal{T}_{f_{n,x}}\Psi\left(0_{j},0_{j}
\right)\mathcal{T}_{f_{n,x}}\Psi\left(0_{j},0_{j}\right)^{\dagger}
\right>_{\mathbb{C}}.\label{F}
\end{equation}
Interestingly, from the CFT point of view, $F^{(n)}_\Psi$ is intrinsically regular, 
since the cutoffs $a$ and $L$ only appear in multiplicative 
state-independent factors in the normalization $\mathcal{N}_n$. 

The EEE is related to the relative R\'enyi entropy of the excited state compared to the vacuum (for some recent general results on relative entropies in 2D CFT see Ref. \cite{Sarosi2016}). The relative entropy is defined as
\begin{equation}
S_n(\rho_\Psi || \rho_\Phi)=\frac{1}{n-1}\left(\log_2\Tr\rho_\Psi^n-\log_2\Tr\rho_\Psi\rho_\Phi^{n-1}\right).
\end{equation}
The $n$-th excess and relative entropies compared to the vacuum (i.e. $\Phi=1$) are related by the following:
\begin{equation}
S_n(\rho_\Psi || \rho_\Phi)=\frac{1}{n-1}\left(\log_2 F_\Psi^{(n)}-\log_2\langle\mathcal{T}_{f_n,x}\Psi(0_1,0_1)\mathcal{T}_{f_n,x}\Psi(0_1,0_1)^\dagger\rangle_\mathbb{C}\right)
\end{equation}.
The appearing two-point functions can be calculated straightforwardly once the transformation laws are known.

The best-known result that can be obtained from Eq. (\ref{path-int}) is the 
following formula for the REE's of the GS of a finite system 
with PBC \cite{Holzhey1994}. Such state is associated with the identity 
operator, for which the correlation function in Eq. (\ref{path-int}) is $1$ and 
only the constant $\mathcal{N}_n$ plays a role. Explicitly, Eq. (\ref{logNn}) looks:
\begin{equation}\label{CCREE}
S_n^{\rm{GS}}(x,L)=\frac{c}{6}\left(1+\frac{1}{n}\right)\log_2\left[\frac{L}{
\pi}\sin\left(\pi x\right)\right]+c_n,
\end{equation}
where $c_n$ is a non-universal constant accounting for regularization. This formula has become, in the years, 
the most used tool in order to extract the central charge of a CFT from 
finite-size numerical data: for most of the available algorithms, REE's are 
among the most natural quantities to compute \cite{Schollwock2011,Humeniuk2012}. Moreover, traditional numerical 
methods need, in order to compute the central charge, the knowledge of 
information about both GS and the first excitations \cite{Blote1986,Affleck1986}, while Eq. 
(\ref{CCREE}) just involves data from the former.

If $\Psi$ is not the identity, corrections can arise, their precise shape 
depending on the considered CFT and the operator itself. Such corrections can be computed using some results from meromorphic CFT \cite{Gaberdiel2000}. First, the adjoint of 
a field can be obtained by 
transforming it according to the map $z\rightarrow z^{-1}$ \cite{DiFrancesco1997}; moreover, a sequence 
of mappings $z\rightarrow f(z)\rightarrow g(f(z))$ has the same effect on any 
operator as the single transformation $z\rightarrow g(f(z))$, since only derivatives, logarithmic 
derivatives and derivatives of the Schwarzian derivative \cite{DiFrancesco1997} occur in 
the transformation laws (see \ref{appA}). Therefore, using the relations
\begin{equation}
\mathcal{T}_{f_{n,x}}\Psi(0)^\dagger=\mathcal{T}_{g_{n,x}}\Psi(0),\qquad g_{n,x}(\xi)=f_{n,x}(1/\xi)=f_{n,-x}(\xi),
\end{equation}
(we supposed the field to be real), the $n$-th EEE reads
\begin{equation}
F^{(n)}_{\Psi}=\left< 
\prod_{j=1}^{n}\mathcal{T}_{f_{n,x}}\Psi(0_j,0_j)\mathcal{T}_{f_{n,-x}}\Psi(0_{j},0_j)\right>_{\mathbb{C}}.
\end{equation}
The operators $\mathcal{T}_{f_{n,x}}\Psi$ can be determined straightforwardly from $\Psi$. The generic transformation rule for secondary fields \cite{Gaberdiel1994} yields a sum of lower descendants in the same tower (see   \ref{appA} for details): the resulting $2n$-point functions of secondary fields are thus evaluated by relating them to $2n$-point functions of primaries.
The two are in general connected by the action of a complicated 
differential operator, which however can be determined (case by case) by means of a systematic approach (see Ref. \cite{Palmai2014} and \ref{appB}). Following this program, the $n$-th EEE finally looks
\begin{equation}
F^{(n)}_{\Psi}=\mathcal{D}_{x,n}^\Psi\left< 
\prod_{j=1}^{n}\Xi(z^+_j,\bar{z}^+_j)\Xi(z^-_j,\bar{z}^-_j)\right>_{\mathbb{C}},
\quad
z^{\pm}_j=e^{\frac{i\pi}{n}(\pm x+2j-2)},
\end{equation}
where $\mathcal{D}_n^\Psi(x)$ is the cited differential operator, and $\Xi$ is 
the primary field in the tower of $\Psi$. In this work, we will need the explicit form of a number of such $2n$-point functions of primary fields. In some cases, we will use, when available, known results from the literature; in the remaining ones, we will compute the correlations by means of the Coulomb-gas approach \cite{DiFrancesco1997}, or different representations of the considered CFT allowing for their derivation.

Finally, we remark that the strategy described in this Section and adopted in Ref. \cite{Palmai2014} is not exactly the one used in Ref. \cite{Berganza2012} for primary states of periodic systems. In particular, we treated the problem starting directly from a replica manifold of planes, instead of ``physical'' cylinders. Descendant states, $\Psi(0)\vert 0\rangle$, are more naturally defined on the plane, where the physical Hamiltonian takes the form $H\sim L_0+\bar L_0$ and descendants are obtained by acting with strings of Virasoro generators on primary states \cite{DiFrancesco1997}. Furthermore, we are allowed to start from planes since a sequence of two conformal mappings using two given holomorphic functions is equivalent to one map under the composition of these functions. Coming back to the physical manifold would be redundant, and more importantly, we observed that starting from the planes simplifies the actual computations substantially by removing uncomfortable infinities, that would need careful regularization. We will use this approach also for the case of OBC.

\subsection{Open boundary conditions}\label{CFT_OBC_sec}

In CFT, OBC reduce the operator content of the theory, and for minimal 
models the OBC preserving conformal invariance are in one-to-one 
correspondence with the primary fields of the theory on the plane 
\cite{Cardy1989}. The partition function of a CFT on the upper half plane, that 
is the prototypical boundary manifold, looks \cite{DiFrancesco1997}
\begin{equation}\label{Z_OBC}
Z_{\alpha\beta}(q)=\sum_h\mathcal{N}^{h}_{\alpha\beta}\chi_h(q),
\end{equation}
with $\alpha$ and $\beta$ being conformal weights of primary fields, indexing 
the possible conformal OBC, $\mathcal{N}^h_{\alpha\beta}$ the fusion 
coefficients, and $\chi_h(q)$ the character corresponding to the primary 
operator of chiral dimension $h$. By expanding Eq. (\ref{Z_OBC}) around 
$q=0$, a series is obtained, whose integer coefficients are the degeneracies of the energy levels. 

For OBC, the physical space-time is an infinite strip of finite width $L$, described, again, by the complex variable $r-i\tau$; the subsystem $A$ is taken as the $\tau=0$ interval $[0,\ell]$. The infinite strip can be mapped to the upper half plane by the transformation $\xi=e^{i\frac{\pi}{L}(r-i\tau)}$ \cite{DiFrancesco1997}; $A$ is then mapped to the unit arc $\mathcal{A}$ connecting $0$ and $e^{i\pi x}$, and the operators associated with the excitation are placed at the origin and at infinity.

Opportunistically, we set up our framework for the REE in the previous subsection in a way that it needs no modification for OBC.
We use the same mapping (\ref{CT}) unifying in this case the $n$ half-planes to a single unit disk (see Figure \ref{transf}).
After the transformation, the operators lie on the boundaries of the disk separating arcs with  different conformal OBC. The resulting $2n$-point functions can be evaluated using boundary CFT: compared to the PBC case now one of the chiralities is suppressed and the chiral building blocks (conformal blocks) combine with different, boundary-dependent, coefficients. This can be understood considering that with OBC present in the system, some fusion channels, open in the PBC case, are now closed and the operator product expansion coefficients \cite{Runkel1999,Runkel2000} determining the weight of the conformal blocks in the $2n$-point functions change. In particular, in the fusion of the boundary operators $\Phi_i^{(\alpha\beta)}(z_i)$ and $\Phi_j^{(\beta\gamma)}(z_j)$, changing the boundary condition at the insertion points from $\alpha$ to $\beta$ and $\beta$ to $\gamma$, respectively, will only involve fields whose towers are present in the partition function $Z_{\alpha\gamma}$ (see Eq. (\ref{Z_OBC})). This fact will be very useful in the situations where we will have to decide which fusion channels are to be considered in order to compute the desired correlators.

\subsection{The identity tower}\label{Tsection}

We now discuss the EEE for states associated to fields in the tower 
of the identity. Such states are present in any bulk CFT, and are also very 
common for theories living on manifolds with boundaries. Moreover, they have the 
property of depending explicitly on the central charge of the CFT: they thus allow, in principle, 
the determination of $c$ from numerical data, generalizing what is usually done 
for the GS to the whole tower.

As an example, we compute the $n=2$ EEE for the first descendant in 
the tower, i.e., the state associated with the stress-energy tensor 
$T=L_{-2}\mathbb{I}$ ($\left\{L_p,\;p\in\mathbb{Z}\right\}$ form the Virasoro 
algebra \cite{DiFrancesco1997}). Its transformation under a conformal map $f$ is
\begin{equation}\label{transf_T}
U_f^{-1}\left(L_{-2}\mathbb{I}\right)(\xi)U_f=\left[f'(\xi)\right]^{2}\left(L_{
-2}\mathbb{I}\right)\left(f(\xi)\right)+\frac{c}{12}\left\{f,\xi\right\}\mathbb{
I},
\end{equation}
being $\{f,\xi\}=f'''(\xi)/f'(\xi)-3\left[f''(\xi)/f'(\xi)\right]^2/2$ the 
Schwarzian derivative of $f$. We can thus write
\begin{equation}\label{L2I4p}
\fl\langle\left(L_{-2}\mathbb{I}\right)(0_1)\left(L_{-2}\mathbb{I}\right)(0_1)^\dagger 
\left(L_{-2}\mathbb{I}\right)(0_2)\left(L_{-2}\mathbb{I}\right)(0_2)^\dagger\rangle_{\mathcal
{R}_2}=\sum_{j,k,l,m=1}^{2}C_{1j}C_{2k}C_{3l}C_{4m}P_{jklm},
\end{equation}
where 
$P_{jklm}=\left<\phi_j(z^+_1)\phi_k(z^-_1)\phi_l(z^+_2)\phi_m(z^-_2)\right>$ is 
the collection of complex-plane four-point functions of the operators $\phi_1=\mathbb{I}$ and 
$\phi_2=L_{-2}\mathbb{I}$ inserted at the appropriate points; the matrix $C$ 
contains the (potentially vanishing) coefficients of the different correlations, and it can be guessed from the transformation relations
\begin{eqnarray}\label{Ttr}
\eqalign{\mathcal{T}_{f_{2,x}}\left(L_{-2}\mathbb{I}\right)(0_{1}) & = 
\sin^2(\pi x)\left[\frac{c}{8}\,\mathbb{I}-e^{i\pi 
x}\left(L_{-2}\mathbb{I}\right)\right]\left(+e^{+i\pi x/2}\right),\\
\mathcal{T}_{f_{2,x}}\left(L_{-2}\mathbb{I}\right)(0_{2}) & =
\sin^2(\pi x)\left[\frac{c}{8}\,\mathbb{I}-e^{i\pi 
x}\left(L_{-2}\mathbb{I}\right)\right]\left(-e^{+i\pi x/2}\right).}
\end{eqnarray}
To obtain $P_{jklm}$, we employ the recipe described in  
\ref{appB} and derive the two-, three- and four-point functions of the 
stress-energy tensor ($T$ is self-adjoint):
\begin{eqnarray}
\fl
\eqalign{\left\langle\left(L_{-2}\mathbb{I}\right)(z_{1})\right\rangle  = 0\cr
\left\langle\left(L_{-2}\mathbb{I}\right)(z_{1})\left(L_{-2}\mathbb{I}\right)(z_{2})\right\rangle  = 
\frac{c}{2z_{12}^{4}},\cr
\left\langle\left(L_{-2}\mathbb{I}\right)(z_{1})\left(L_{-2}\mathbb{I}\right)(z_{2})\left(L_{-2}\mathbb{I}\right)(z_{3}
)\right\rangle  = 
\frac{c}{z_{12}^{2}z_{13}^{2}z_{23}^{2}},\cr
\left\langle\left(L_{-2}\mathbb{I}\right)(z_{1})\left(L_{-2}\mathbb{I}\right)(z_{2})\left(L_{-2}\mathbb{I}\right)(z_{3})\left(L_{-2}
\mathbb{I}\right)(z_{4})\right\rangle = 
\sum_{jkl=234,324,423}\frac{c^{2}/4}{z_{1j}^{4}z_{kl}^{4}}+\cr
  \qquad\qquad\qquad\qquad+2c\frac{\sum_{j<k}z_{j}^{2}z_{k}^{2}-\sum_{j}\sum_{k<l\in\left\{ 
1,2,3,4\right\} \setminus 
j}z_{j}^{2}z_{k}z_{l}+6z_{1}z_{2}z_{3}z_{4}}{\sum_{j<k}z_{jk}^{2}}.}
\end{eqnarray}
being $z_{jk}=z_j-z_k$. After substituting $z^\pm_{1,2}$, performing the sums in Eq. (\ref{L2I4p}) with 
the appropriate coefficients $C$, determined from (\ref{Ttr}), and multiplying 
by the normalization $\sqrt{c/2}$ for the state $\vert 
L_{-2}\mathbb{I}\rangle$, we obtain
\begin{equation}\label{F_T}
\fl\eqalign{F_{T}^{(2)}(x)  = &\frac{\sin^{4}(\pi x)\left[\cos(2\pi 
x)+7\right]}{16}c^{-1}+ \cr
&+\frac{16200\cos(2\pi x)-228\cos(4\pi x)+120\cos(6\pi 
x)+\cos(8\pi x)+16675}{32768}+ \cr
 &  +\frac{\sin^{4}(\pi x)\left[\cos(2\pi 
x)+7\right]^{2}}{1024}c+\frac{\sin^{8}(\pi x)}{1024}c^{2},}
\end{equation}
that is what we will compare to numerical data in the next Sections.

\begin{figure}[t]
\centering
\includegraphics[width=\linewidth,trim=0cm 10.1cm 2cm 0.2cm, clip=true]{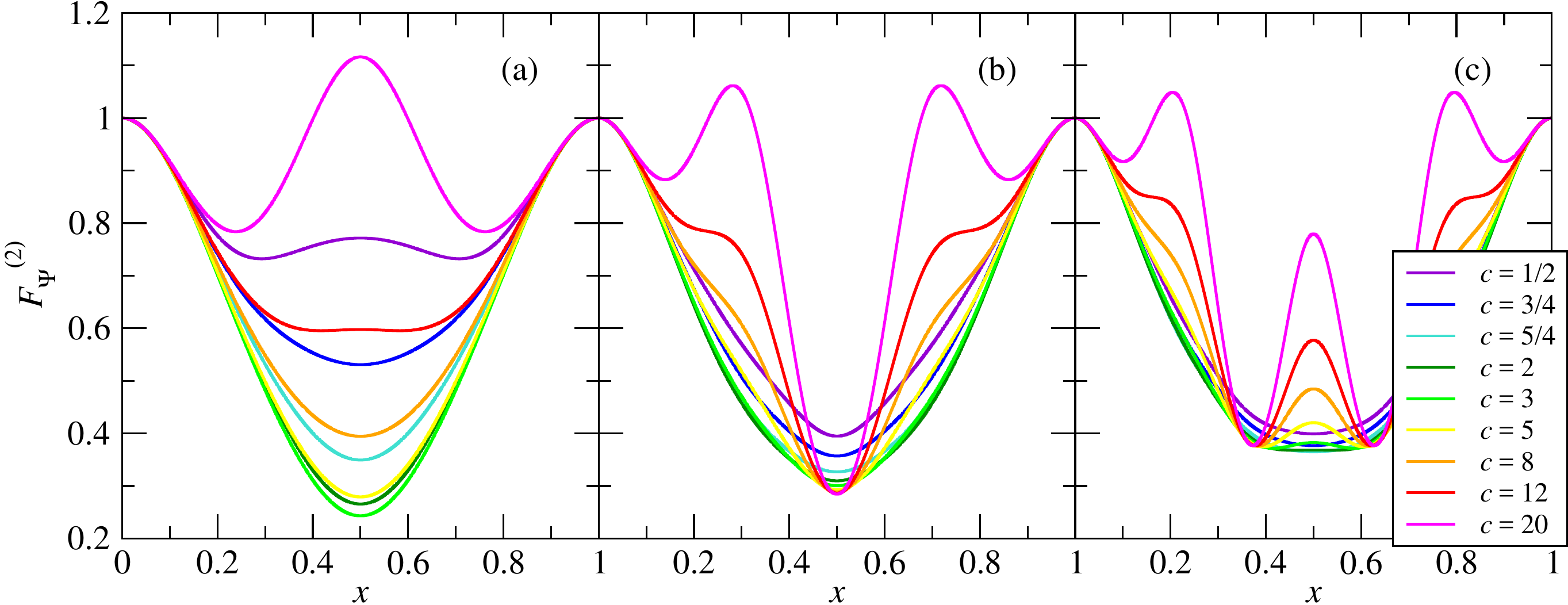}
\caption{$F^{(2)}_\Psi(x)$ for $\Psi=L_{-2}$ (panel (a)), $L_{-3}$ (panel (b)) and $L_{-4}$ (panel (c)), for several values of the central charge $c$. }\label{figT}
\end{figure}

Before doing it, it is worth to analyze Eq. (\ref{F_T}) both as a function of the relative subsystem size $x$ and of the central charge $c$. In Fig. \ref{figT}(a) we show $F_T^{(2)}(x)$ for different values of $c$. We observe that the small-block EEE is independent of $c$, which can also be seen by expanding $F_T^{(2)}(x)$ around $x=0$:
\begin{equation}\fl
F_T^{(2)}(x)=1-(\pi x)^2+\frac{3c^2+22c+24}{48c}(\pi x)^4-\frac{105 c^2+364 c+660}{1440 c}(\pi x)^6+O(x^8).\end{equation}
This behavior is in agreement with the holographic result of Ref. \cite{Bhattacharya2013}, where it was established that the excess VNEE is proportional to the excitation energy, i.e., $\Delta S=\frac{1}{3}\pi x\Delta E$, and in particular it is independent of $c$. In the opposite limit, the half-block EEE is given by
\begin{equation}\fl
F_T^{(2)}(1/2)=\frac{1}{256}\left(\frac{c^2}{4}+9c+1+\frac{3}{32c}\right).
\end{equation}
Interestingly, this function features a minimum located at $c_{\rm min}\approx 3.02212$. For small $c$, the whole function $F_T^{(2)}(x)$ diverges, signaling that in a $c=0$ unitary CFT only the vacuum state exists \cite{Gomes1986}. When $c$ is large, we see from the curves of Fig. \ref{figT}(a) that the leading term $c^2\sin^8(\pi x)/1024$ starts to dominate. This leading term emerges when taking only the contributions of the identity in the transformation laws (\ref{transf_T}). This is the regime where the AdS/CFT correspondence should play a role (see, for a review, e.g., Ref. \cite{Nishioka2009}); however, to our knowledge, this result is not yet available in that context (again, we remark that results for the small-block limit are already available \cite{Bhattacharya2013}.)

While in the present work we will only use the result for the stress-energy tensor, we emphasize that by means of the general transformation rule described in \ref{appA} a similar analysis can be carried out for the whole identity tower. E.g., for the state related to $L_{-3}\mathbb{I}$ we obtain
\begin{equation}\label{F_L3}
\fl\eqalign{F_{L_{-3}\mathbb{I}}^{(2)}  = &\frac{\sin ^4(\pi  x) \left[8391 \cos (2 \pi  x)+1890 \cos (4 \pi  x)+361 \cos (6 \pi  x)+7790\right]}{16384}c^{-1}+ \cr
&+\frac{3032808 \cos (2 \pi  x)+819919 \cos (4 \pi  x)-27612 \cos (6 \pi  x)}{8388608}+ \cr
&+\frac{386 \cos (8 \pi  x)+8436 \cos (10 \pi  x)+289 \cos (12 \pi  x)+4554382}{8388608}+\cr
 &  +\frac{\sin ^4(\pi  x) \cos ^2(\pi  x) \left[255 \cos (2 \pi  x)+90 \cos (4 \pi  x)+17 \cos (6 \pi  x)+1686\right]}{8192}c\cr
 &+\frac{ \sin ^8(\pi  x) \cos ^4(\pi  x)}{64}c^{2}.}
\end{equation}
In Figs. \ref{figT}(b) and (c) we show the $n=2$ EEE's for $L_{-3}\vert 0\rangle$ and $L_{-4}\vert 0\rangle$ and several values of the central charge. Similarly to the case of $L_{-2}\vert 0\rangle$, the higher descendants also show a $c$-independence for small blocks, in agreement with Ref. \cite{Bhattacharya2013}. 
Comparing the three panels of Fig. 2, we observe that, increasing the level of the excitation, the shape of the curves for $c\lesssim 3$ (these values of course include minimal models) becomes less and less dependent on the actual value of the central charge.


\section{$c=1/2$ minimal theory and spin-1/2 Ising chain in a transverse 
field}\label{Ising_sec}

In this Section, we present analytical predictions for the $c=1/2$ minimal CFT, 
and numerical data relative to the $n=2$ REE for the descendant states of the 
spin-1/2 Ising chain in a transverse field. The REE's for the primary states 
were already discussed in Ref. \cite{Taddia2013}; here, we just 
focus on descendant states. 

The $c=1/2$ minimal model is one of the simplest CFT's \cite{DiFrancesco1997}. The 
operator content of the model on the plane is the following: the primary fields 
are the identity and the fields $\sigma$ and $\psi$, of 
chiral dimension $0$, $1/16$, and $1/2$, respectively. For minimal models on the 
upper half plane, the BC preserving conformal invariance are in 
one-to-one correspondence with such primary fields, leading, as we shall see, to 
four possible pairs of OBC. In this Section we will consider, in any case, 
the first descendant state in each conformal tower.

A 1D lattice realization of the $c=1/2$ minimal CFT is the spin-1/2 Ising chain in 
a transverse field at the critical point \cite{Mussardo2010}. The Hamiltonian of such chain is given by
\begin{equation}\label{Ising_H}
H=-J\sum_{j}\sigma_j^z\sigma_{j+1}^z-h\sum_{j}\sigma_j^x,
\end{equation}
where $\sigma^\alpha_j$ is a Pauli matrix ($\alpha=x,y,z$) at the
site $j\in\{1,2,\ldots,L\}$; the critical point is located at $h/J=1$. The three possible 
conformal OBC are the following \cite{Cardy1986,Zhou2006}: the $x$-component of 
the boundary spins must be fixed to $1/2$ ($+$), $-1/2$ ($-$) or let free ($F$), 
in correspondence, respectively, with the $\mathbb{I}$, $\psi$ and 
$\sigma$ primary fields. Once combined, because of the $Z_2$ symmetry of 
Hamiltonian (\ref{Ising_H}), there are just four independent situations, namely 
$++$, $+-$, $+F$ and $FF$, where the first symbol indicates the boundary 
condition chosen for the first spin and the second for the last. 

For $FF$BC, the Hamiltonian can be written in terms of free spinless fermions by 
means of a Jordan-Wigner transformation \cite{Sachdev2011}. In such case, it can 
be diagonalized in an exact way, exploiting the properties of free fermions; 
consequently, the REE's can be computed exactly, following the recipe of Refs. 
\cite{Vidal2003,Peschel2003}. In the remaining cases, because of the 
presence of the fixed OBC, the Jordan-Wigner-transformed Hamiltonian is 
not anymore quadratic at the boundaries, and the use of approximated techniques is a 
necessity. We perform the computations by means of the density-matrix 
renormalization group (DMRG) technique \cite{White1992} in its multi-target 
version \cite{DegliEspostiBoschi2004}: it allows for a straightforward 
computation of the REE's of the first excited states of the chain, as well as 
the implementation of the OBC in an exact way \cite{Taddia2014}. In any case, we 
consider systems up to $L=1000$ sites; in the DMRG calculations, we employ 7 
finite-size sweeps and keep up to 64 states, achieving, in the last steps, a 
maximum truncation error of the order of $10^{-8}$.

What stated in this paragraph also holds for the results of Sections \ref{Potts} 
and \ref{XX}. The obtained numerical data has, in any case, been tested in two 
independent ways: by comparing the numerical degeneracies of energy multiplets 
and the ones predicted by CFT (see, e.g., Eq. (\ref{ZIsing})); by computing the 
conformal weight of the considered states from the finite-size scaling (FSS) of 
their energies, according to the CFT formula \cite{Blote1986,Affleck1986}
\begin{equation}
E_h(L)-E_0(L)=\frac{\pi u}{L}h,
\end{equation}
being $E_h(L)$ the energy of the state of weight $h$ at size $L$, and $u$ the 
sound velocity of the system, known to be $1$ for the spin-1/2 Ising chain in a 
transverse field and the spin-1/2 XX chain \cite{Taddia2014}. For the 
three-state Potts chain, the sound velocity has been extracted numerically from 
the finite-size scaling of the GS energy density with $(A,A)$ boundary 
conditions (see Section \ref{Potts}), i.e., from 
\cite{Blote1986,Affleck1986,Taddia2014}
\begin{equation}
e_0(L)=e_b+\frac{e_s}{L}+\frac{uc\pi}{24L^2}+o\left(L^{-2}\right),
\end{equation}
where $e_b$ and $e_s$ are, respectively, the bulk and surface component of the 
energy density. For the three-state Potts chain, we obtain $u=5/2$ (not 
shown).

To compare our results for model (\ref{Ising_H}) with the CFT predictions we need to identify 
the low-energy states in the two frameworks. As pointed out by Cardy 
\cite{Cardy1986}, the operator content of the low-energy effective field theory 
is affected by the BC in the following way:
\begin{eqnarray}\label{Z_Ising}
Z_{++}(q) & = \chi_{0}(q),\\
Z_{+-}(q) & = \chi_{1/2}(q),\\
Z_{+F}(q) & = \chi_{1/16}(q),\\
Z_{FF}(q) & = \chi_{0}(q)+\chi_{1/2}(q).
\end{eqnarray}
We can expand these partition functions in powers of $q$ in order to determine
the degeneracies of the excited states \cite{DiFrancesco1997}:
\begin{equation}\label{ZIsing}
\eqalign{q^{c/24}Z_{++}(q) & = 1+q^2+\mathcal{O}\left(q^3\right),\cr
q^{c/24}Z_{+-}(q) & = q^{1/2}+q^{3/2}+\mathcal{O}\left(q^{5/2}\right),\cr
q^{c/24}Z_{+F}(q) & = q^{1/16}+q^{17/16}+\mathcal{O}\left(q^{33/16}\right),\cr
q^{c/24}Z_{FF}(q) & = 1+q^{1/2}+q^{3/2}+q^2+\mathcal{O}\left(q^{5/2}\right).}
\end{equation}
The first descendant states are, in any case, non degenerate: this makes 
our task easier, since, in general, the DMRG algorithm, when dealing with degeneracies, 
considers a non-trivial linear combination in the multiplet. The analysis of such case will be 
unavoidable when we will study, in Section \ref{XX}, the spin-1/2 XX chain.

\begin{figure}[h!]
\centering
\includegraphics[width=0.9\linewidth,trim=0cm 0cm 0cm 3.7cm, clip=true]{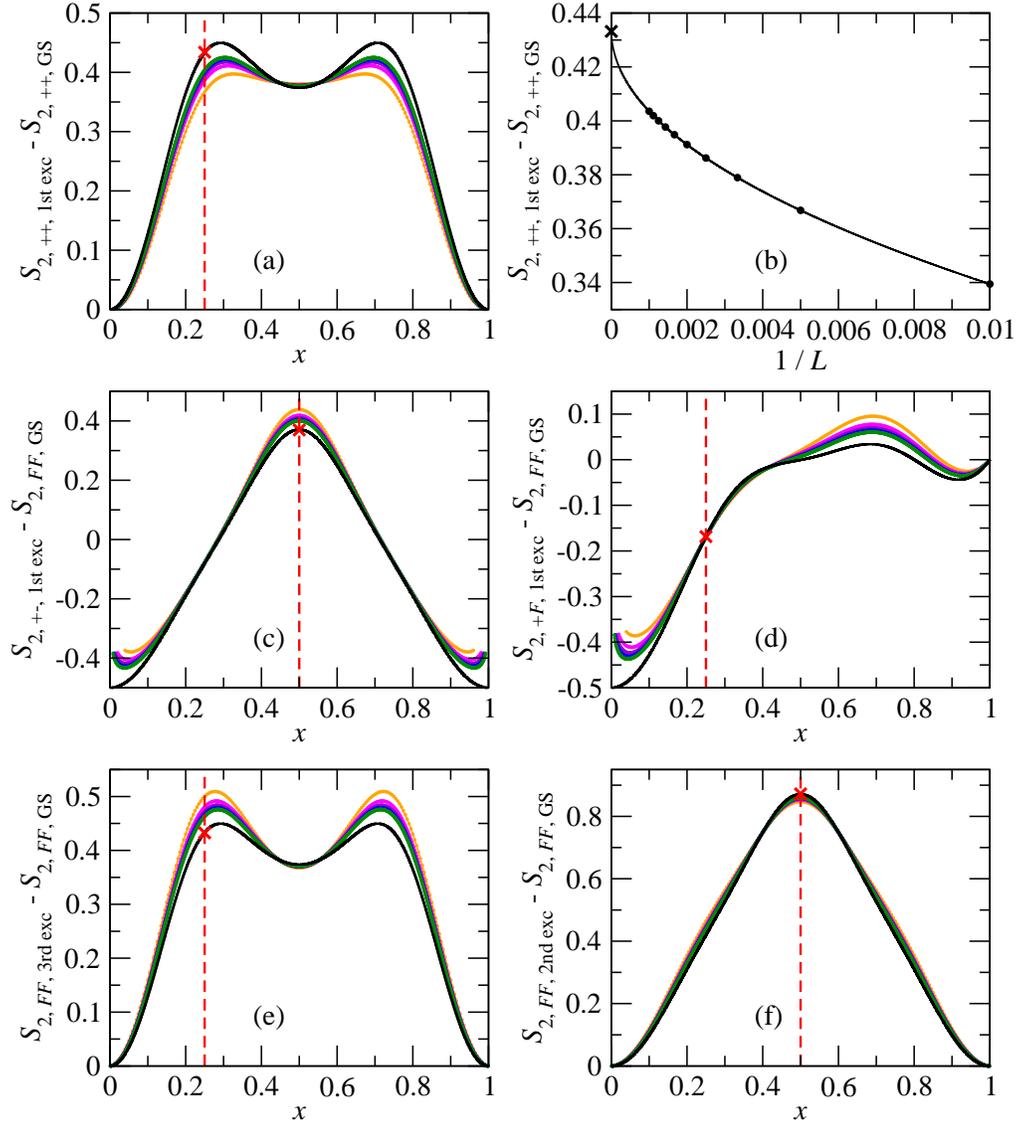}
\caption{$n=2$ EEE for the first descendant states of the critical 
spin-1/2 Ising chain in a transverse field with conformal OBC. Panels (a), 
(c)-(f): difference between $S_2$ for the considered descendant state and OBC 
(see Table \ref{table:ising_fit} for details) and the GS (for some 
relevant conformal OBC) as a function of $x$ and for different values of $L$. 
Black line: CFT prediction; colored dotted lines: DMRG data (orange to green: 
$L=200$ to $1000$); vertical red dashed line: considered $x$ for the FSS analysis (see 
Table \ref{table:ising_fit} for details); red cross: result of the FSS analysis. Panel 
(b): FSS of the data of panel (a) at $x=1/4$. Circles: DMRG data; cross: CFT 
prediction; solid line: best fit, by means of the three-parameters formula 
$y=a_0+a_1x^{a_2}$ (see Table \ref{table:ising_fit} for the obtained values of the 
fir parameters and the deviations from the CFT 
predictions).}\label{Ising_fig}
\end{figure}

We start the analysis with the $++$ case, where the only present tower is the one of the 
identity. Therefore, the first descendant state is the one associated with the 
stress-energy tensor, and its $n=2$ EEE takes the form in Eq. 
(\ref{F_T}), with $c=1/2$:
\begin{equation}\label{F_T_Ising}
\fl F_{T}^{(2)}(x)=\frac{426347+53640\cos(2\pi x)+38076\cos(4\pi x)+6200\cos(6\pi 
x)+25\cos(8\pi x)}{524288}.
\end{equation}
We show, in Figure \ref{Ising_fig}(a), the difference between the $n=2$ REE for the 
first excited state and for the GS: as shown in Figure 
\ref{Ising_fig}(b) and Table \ref{table:ising_fit}, a FSS analysis confirms that, in the 
thermodynamic limit, the DMRG data approaches the CFT prediction with great 
accuracy.
\begin{table}[t]
 \centering
 \begin{tabular}{|c|c|c|c|c|c|c|c|c|}
  \hline
  BC & $m$ & $h$ & $x$ & $a_0$ & $a_1$ & $a_2$ & $\eta$ & Figure\\
  \hline
  $++$ & $1$ & $2$ & $1/4$ & $0.434$ & $-0.915$ & $0.493$ & $1.6\times10^{-3}$ & 
\ref{Ising_fig}(a), (b)\\
  \hline
  $+-$ & $1$ & $3/2$ & $1/2$ & $0.370$ & $0.910$ & $0.487$ & $3\times10^{-3}$ & 
\ref{Ising_fig}(c)\\
  \hline
  $+F$ & $1$ & $17/16$ & $1/4$ & $-0.169$ & $-0.965$ & $1.216$ & 
$2\times10^{-3}$ & \ref{Ising_fig}(d)\\
  \hline
  $FF$ & $3$ & $2$ & $1/4$ & $0.433$ & $0.921$ & $0.494$ & $1.4\times10^{-3}$ & 
\ref{Ising_fig}(e)\\
  \hline
  $FF$ & $2$ & $3/2$ & $1/2$ & $0.872$ & $-0.265$ & $0.444$ & $1.7\times10^{-3}$ 
& \ref{Ising_fig}(f)\\
  \hline
 \end{tabular}
 \caption{Fitting result of the finite-size data for the transverse-field Ising 
chain with OBC. First column: considered BC; second column: 
position in the energy spectrum (e.g., $m=1$: first excited state); third 
column: conformal weight of the considered state; fourth column: value of $x$ at 
which the finite-size fite is performed; fifth, sixth and seventh columns: 
estimated fit parameters (by means of the formula: $y=a_0+a_1x^{a_2}$); eighth column: 
relative deviation $\eta$ of $a_0$ from the CFT prediction; nineth column: 
corresponding Figure.}\label{table:ising_fit}
\end{table}
Such non-oscillating (compared to the $c=1$ case; see 
Section \ref{XX}) finite-size behavior is typical of spin-1/2 XY chains \cite{Igloi2008,Taddia2013}, 
and will be also observed for the three-state Potts chain in Section 
\ref{Potts}, thus suggesting a general behavior for minimal CFT's.

We next consider the case of $+-$BC. The operator content is 
the tower of the $\psi$ operator, and the first secondary operator in the tower 
is $L_{-1}\psi$. The relevant four-point function is easily obtained by the Wick 
theorem \cite{DiFrancesco1997} ($\psi$ is a free Fermi field \cite{Mussardo2010}):
\begin{equation}
\langle\psi(z_1)\psi(z_2)\psi(z_3)\psi(z_4)\rangle=\sum_{jklm=1234,1423,2431}
\frac{1}{z_{jk}z_{lm}},
\end{equation}
Using the relation above and the transformation rule (\ref{eq:L-1}) we find, with a procedure 
similar to the one used for the first descendant in the identity tower in Section 
\ref{Tsection},
\begin{equation}\label{F_+-}
F_{L_{-1}\psi}^{(2)}(x)=\frac{1558+439\cos(2\pi x)+26\cos(4\pi x)+25\cos(6\pi 
x)}{2048}.
\end{equation}
We show in Figure \ref{Ising_fig}(c) the difference between the $n=2$ REE for the 
first excited state and for the GS with $FF$BC (for 
such BC the GS is associated to the identity 
operator, and therefore taking the difference just the correction due to the 
secondary operator shall survive, at least apart from finite-size corrections). Also 
in this case, the agreement between the CFT prediction and the numerical data 
scaled to the thermodynamic limit is excellent (see table \ref{table:ising_fit} 
for details).
We note that the formalism we have developed is, alone, not able to fully 
capture the numerical behaviour of the EEE. Indeed, we find that, in order to 
reproduce it, we have to support Eq. (\ref{F_+-}) with the additive 
contribution 
\begin{equation}
S_b=\log_2g,
\end{equation}
that takes, in the present case, the value $-1/2$ (i.e., $g=1/\sqrt{2}$): it can be interpreted as the 
Affleck-Ludwig boundary REE \cite{Affleck1991} associated with the considered 
BC. The fact that the difference with the entropy of the GS with $FF$BC is considered is crucial: indeed, Zhou and 
collaborators showed, in Ref. \cite{Zhou2006} and for the GS's, 
that $g$ takes the value $1/\sqrt{2}$ for $\pm$BC, and $0$ for 
$F$. The same situation, i.e., that the boundary entropy has to be added to the 
CFT prediction, was observed for primary excited states in Ref. \cite{Taddia2013}.

The third case we consider is the one of $+F$BC, leading to 
a theory containing only the tower of the $\sigma$ field; the first 
secondary operator in the theory is thus $L_{-1}\sigma$. The four-point function that is relevant for the present case reads \cite{Ardonne2010}:
\begin{equation}\label{sigma4}
\fl\eqalign{\left<\sigma(z_1)^{(+F)}\sigma(z_2)^{(F+)}\sigma(z_3)^{(+F)}\sigma(z_4)^{(F+)}\right>_\mathbb{C}=\left(\frac{z_{13}z_{24}}{z_{12}z_{14}z_{23}z_{34}}\right)^{1/8}\sqrt{1+\sqrt{\frac{z_{12}z_{34}}{z_{13}z_{24}}}}},
\end{equation}
which is nothing but the conformal block corresponding to the identity channel 
in the PBC fusion $\sigma\times\sigma=\mathbb{I}+\epsilon$ \cite{DiFrancesco1997}. 
This is the correct four-point function to be considered, since, as argued at the end of Section \ref{CFT_OBC_sec}, only the fusion channels allowed by the partition function $Z_{++}$ must be considered (see Eq. (\ref{Z_Ising})). 
Using Eq. (\ref{sigma4}) and the transformation rule (\ref{eq:L-1}) for $L_{-1}\sigma$, the $n=2$ EEE can be seen to take the form: 
\begin{equation}\label{F_+F}
\fl \eqalign{F_{L_{-1}\sigma}^{(2)} =\frac{1}{512\sqrt{2}} \left(\cos\left(\frac{\pi 
x}{4}\right)+\sin\left(\frac{\pi 
x}{4}\right)\right)\Biggl[-16\sin\left(\frac{\pi 
x}{2}\right)+16\sin\left(\frac{3\pi x}{2}\right)\Biggr. \cr +\Biggl.16\sin\left(\frac{5\pi 
x}{2}\right)-16\sin\left(\frac{7\pi x}{2}\right)
-16\cos(\pi x)+60\cos(2\pi x)+16\cos(3\pi x)+9\cos(4\pi x)+443\Biggr].}
\end{equation}
Again, we consider, numerically, the first excited state, and we show its $n=2$ EEE in 
Figure \ref{Ising_fig}(d): now, the profile is highly asymmetric with respect to 
$x=1/2$, because of the different OBC at the boundaries, and interpolates 
non-trivially between $-1/2$ and 0. Again a boundary entropy $-1/2$ has to be 
added to the CFT prediction in order for it to match the numerical data, and a 
FSS analysis confirms the correctness of the analytical approach.

The last case we consider is the $FF$ case, for which the theory contains two 
characters, relative to the identity and to the $\psi$ operator. The first 
descendant states in each tower are, respectively, the third and the second 
excited states of the chain, corresponding to the stress-energy tensor and to  
$L_{-1}\psi$; the corrections for them have already been derived and are given 
by Eqs. (\ref{F_+-}) and (\ref{F_T_Ising}). The $n=2$ EEE's, obtained 
using exact diagonalization, are plotted in Figs. \ref{Ising_fig}(e) and (f). In 
both cases, the agreement between CFT and the FSS-scaled numerical data is 
remarkable.

To conclude, we were able to interpret the numerical data for the $n=2$ EEE's in all the considered situations, finding excellent quantitative 
agreement with the analytical CFT predictions.

\section{$c=4/5$ minimal CFT and three-state Potts chain}\label{Potts}

We consider, in this Section, the minimal CFT with central charge $c=4/5$ 
\cite{DiFrancesco1997}. The operator content of the theory is richer 
than the one of the $c=1/2$ minimal CFT: there are eight primary fields (with 
respect to an extended $W$-algebra), of conformal dimensions $0$, $1/40$, 
$1/15$, $1/8$, $2/5$, $21/40$, $2/3$ and $13/8$.

A lattice realization of the CFT is the three-state Potts chain at its 
critical point \cite{Wu1982}. It is characterized by the Hamiltonian
\begin{equation}\label{Potts_H}
H=-h\sum_j\left(M_j+M_j^\dagger\right)-J\sum_j\left(R_j^\dagger 
R_{j+1}+R_{j+1}^\dagger R_j\right),
\end{equation}
where $j\in\{1,2,\ldots,L\}$ indexes the lattice site, and
\begin{equation}
M_j=\left(\begin{array}{ccc}
0 & 1 & 0\\
0 & 0 & 1\\
1 & 0 &0
\end{array}\right)_j,\;\;\;R_j=\left(\begin{array}{ccc}
e^{i2\pi/3} & 0 & 0\\
0 & e^{i4\pi/3} & 0\\
0 & 0 & 1
\end{array}\right)_j.
\end{equation}
The critical point is at $h/J=1$, realizing the $c=4/5$ minimal 
theory described above. The conformal OBC have been derived by Cardy 
\cite{Cardy1989} and by Affleck and collaborators \cite{Affleck1998}: they are 
eight, in one-to-one correspondence with the primary fields of the bulk theory. 
According to the matrix notation used for Hamiltonian (\ref{Potts_H}), the first 
three of them, that we will call $A$, $B$ and $C$, consist in fixing the 
boundary state to $\left|A\right>=$(1,0,0$)^T$, $\left|B\right>=$(0,1,0$)^T$ and 
$\left|C\right>=$(0,0,1$)^T$; the second three, that we will call $AB$, $AC$ and 
$BC$, consist in fixing them to 
$\left|AB\right>=\left(\left|A\right>+\left|B\right>\right)/\sqrt{2}$, 
$\left|AC\right>=\left(\left|A\right>+\left|C\right>\right)/\sqrt{2}$ and 
$\left|BC\right>=\left(\left|B\right>+\left|C\right>\right)/\sqrt{2}$; the $F$BC consists in leaving the boundary spin free, and the $N$ in 
adding to the Hamiltonian a term, e.g., on the first site, of the form 
$-h_1(M_1+M_1^\dagger)$, with $h_1\ll0$ \cite{Affleck1998}. When put together in 
couples, because of the $Z_3$ symmetry of Eq. (\ref{Potts_H}), there are fourteen possible choices of conformal OBC: 
$(A,A)$, $(A,B)$, $(A,AB)$, $(A,AC)$, $(A,BC)$, $(A,F)$, $(A,N)$, $(AB,AB)$, 
$(AB,AC)$, $(AB,F)$, $(AB,N)$, $(F,F)$, $(F,N)$ and $(N,N)$. We will consider, 
in the present work, just the three cases $(A,A)$, $(A,B)$ and $(A,F)$.

The corresponding partition functions are given by \cite{Cardy1989,Affleck1998}:
\begin{eqnarray}
Z_{(A,A)}(q)&=\chi_{0}(q),\label{PottsAA}\\
Z_{(A,B)}(q)&=\chi_{2/3}(q),\\
Z_{(A,F)}(q)&=\chi_{1/8}(q)+\chi_{13/8}(q).\label{PottsAfree}
\end{eqnarray}
We note that, while in the PBC case the operator content of the three-state 
Potts chain is reduced to the primary fields of conformal weights $0$, $1/15$, 
$2/5$ and $2/3$ \cite{DiFrancesco1997}, by putting on the edges conformal OBC 
as, e.g., $(A,F)$, it is possible to introduce in the model unusual primary 
fields, occuring in the diagonal theory only \cite{Mussardo2010}. 
By expanding the partition functions around $q=0$, we obtain:
\begin{eqnarray}
\eqalign{q^{c/24}Z_{(A,A)}&(q)=1+q^2+\mathcal{O}\left(q^3\right),\\
q^{c/24}Z_{(A,B)}&(q)=q^{2/3}+q^{5/3}+\mathcal{O}\left(q^2\right),\\
q^{c/24}Z_{(A,F)}&(q)=q^{1/8}+q^{9/8}+q^{13/8}+q^{17/8}+q^{21/8}+\mathcal{O}
\left(q^{25/8}\right).}
\end{eqnarray}
In all of the considered cases, the first descendant states are non-degenerate; 
we will consider, in the towers associated to $h=0$, $1/8$, and $2/3$, the 
first descendant states. In addition, we will also consider the $n=2$ EEE's of primary states, since their analysis is absent in the 
literature. The numerical data is obtained by means of the DMRG algorithm, 
with system sizes up to $L=300$, using up to 7 finite-size sweeps, keeping up to 
200 states and achieving a truncation error at the final stages of the 
finite-system procedure of $10^{-8}$ or less.

\begin{figure}[t]
\centering
\includegraphics[width=\linewidth,trim=0.1cm 0.2cm 1.8cm 0.1cm, 
clip=true]{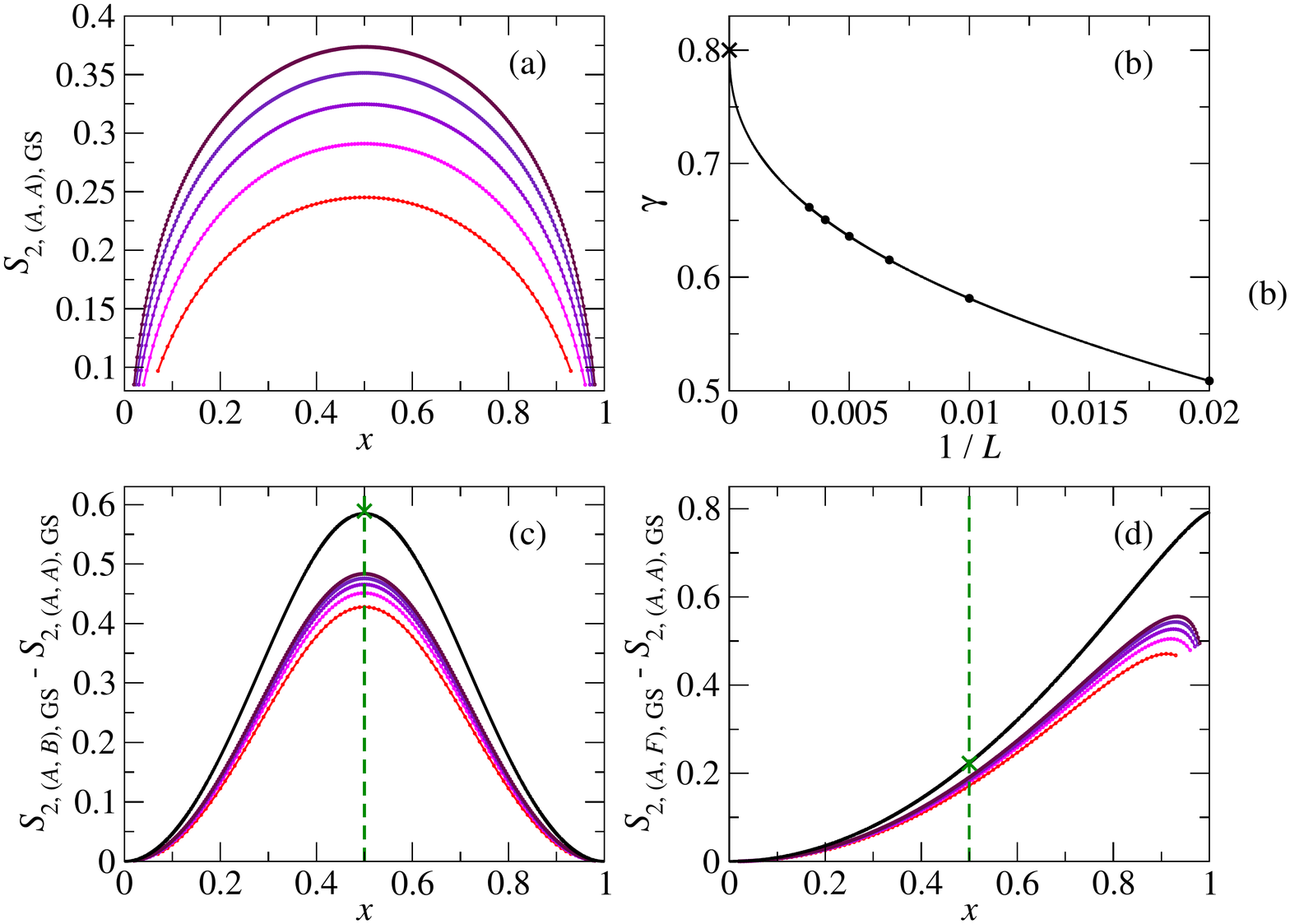}
\caption{$n=2$ REE for primary states of the critical three-state Potts chain 
with conformal OBC. Panel (a): $S_2$ for the GS with $(A,A)$BC at 
different system sizes. Panels (c) and (d): difference between $S_2$ for the 
considered primary state and OBC (see Table \ref{table:PottsPrimaries_fit} for 
details) and the GS with $(A,A)$BC as a function of $x$ and for 
different values of $L$ (see Table \ref{table:PottsPrimaries_fit} for details). 
Black line: CFT prediction; colored dotted lines: DMRG data (red to maroon: 
$L=100$ to $300$); vertical green dashed line: considered $x$($=1/2$) for the 
FSS analysis; green cross: result of the FSS analysis. Panel (b): FSS of the fitted value of the 
central charge $\gamma(L)$, as obtained from the data in panel (a). Circles: DMRG data; 
cross: CFT prediction; solid line: best fit, by means of the 
three-parameters formula $y=a_0+a_1x^{a_2}$ (see Table 
\ref{table:PottsPrimaries_fit} for the obtained values of the coefficients and the deviations from the CFT predictions).}\label{PottsPrimaries_fig}
\end{figure}

We start by considering the $(A,A)$ case: the operator content is very simple, 
since the partition function only contains the character of the identity. The 
GS is associated with the identity itself, and the $n$-th REE's must 
follow the scaling \cite{Calabrese2004} 
\begin{equation}\label{CCREEOBC}
S_n^{\rm{GS}}(x,L)=\frac{c}{12}\left(1+\frac{1}{n}\right)\log_2\left[\frac{2L}
{\pi}\sin\left(\pi x\right)\right]+d_1,
\end{equation}
slightly different from Eq. (\ref{CCREE}), because of the presence of OBC. Such 
scaling has been verified numerically, by comparing it to the DMRG data, as 
shown in Figure \ref{PottsPrimaries_fig}(a): each of the curves in this Figure, 
corresponding to different values of $L$, can be fitted with good precision by means of Eq. 
(\ref{CCREEOBC}), but the obtained value of the central charge, $\gamma(L)$, suffers from a 
strong finite-size effect, as shown in Figure \ref{PottsPrimaries_fig}(b).
However, in the same Figure, a simple FSS analysis is performed, in order to 
show that, in the thermodynamic limit, such values converge to the theoretical 
one, $c=4/5$ (see the caption of Figure \ref{PottsPrimaries_fig} for the details 
of the FSS analysis, and Table \ref{table:PottsPrimaries_fit} for its 
quantitative results).
\begin{table}[t]
 \centering
 \begin{tabular}{|c|c|c|c|c|c|c|c|c|}
  \hline
  BC & $h$ & $a_0$ & $a_1$ & $a_2$ & $\eta$ & Figure\\
  \hline
  $(A,A)$ & $0$ & $0.803$ & $-1.458$ & $0.409$ & $4\times10^{-3}$ & 
\ref{PottsPrimaries_fig}(a), (b)\\
  \hline
  $(A,B)$ & $2/3$ & $0.870$ & $0.262$ & $0.399$ & $1.0\times10^{-3}$ & 
\ref{PottsPrimaries_fig}(c)\\
  \hline
  $(A,F)$ & $1/8$ & $0.222$ & $-0.370$ & $0.431$ & $3\times10^{-3}$ & 
\ref{PottsPrimaries_fig}(d)\\
  \hline
 \end{tabular}
 \caption{Fitting result of the finite-size data for the GS of the 
three-state Potts chain with OBC. First column: considered BC; 
second column: conformal weight of the considered state; third, fourth and fifth 
columns: estimated fit parameters (used formula: $y=a_0+a_1x^{a_2}$); sixth 
column: relative deviation $\eta$ of $a_0$ from the CFT prediction; seventh 
column: corresponding Figure. The fitted quantity is different in the first and 
the remaining two cases; see the main text for details about the procedures. In 
any case, the analysis is performed at a subsystem size 
$x=1/2$.}\label{table:PottsPrimaries_fit}
\end{table}
 We were therefore able to show that the $n=2$ REE of the GS displays 
a behavior that is compatible with the scaling (\ref{CCREEOBC}).

\begin{figure}[ht]
\centering
\includegraphics[width=\linewidth,trim=0.4cm 0.2cm 1.8cm 0.4cm, 
clip=true]{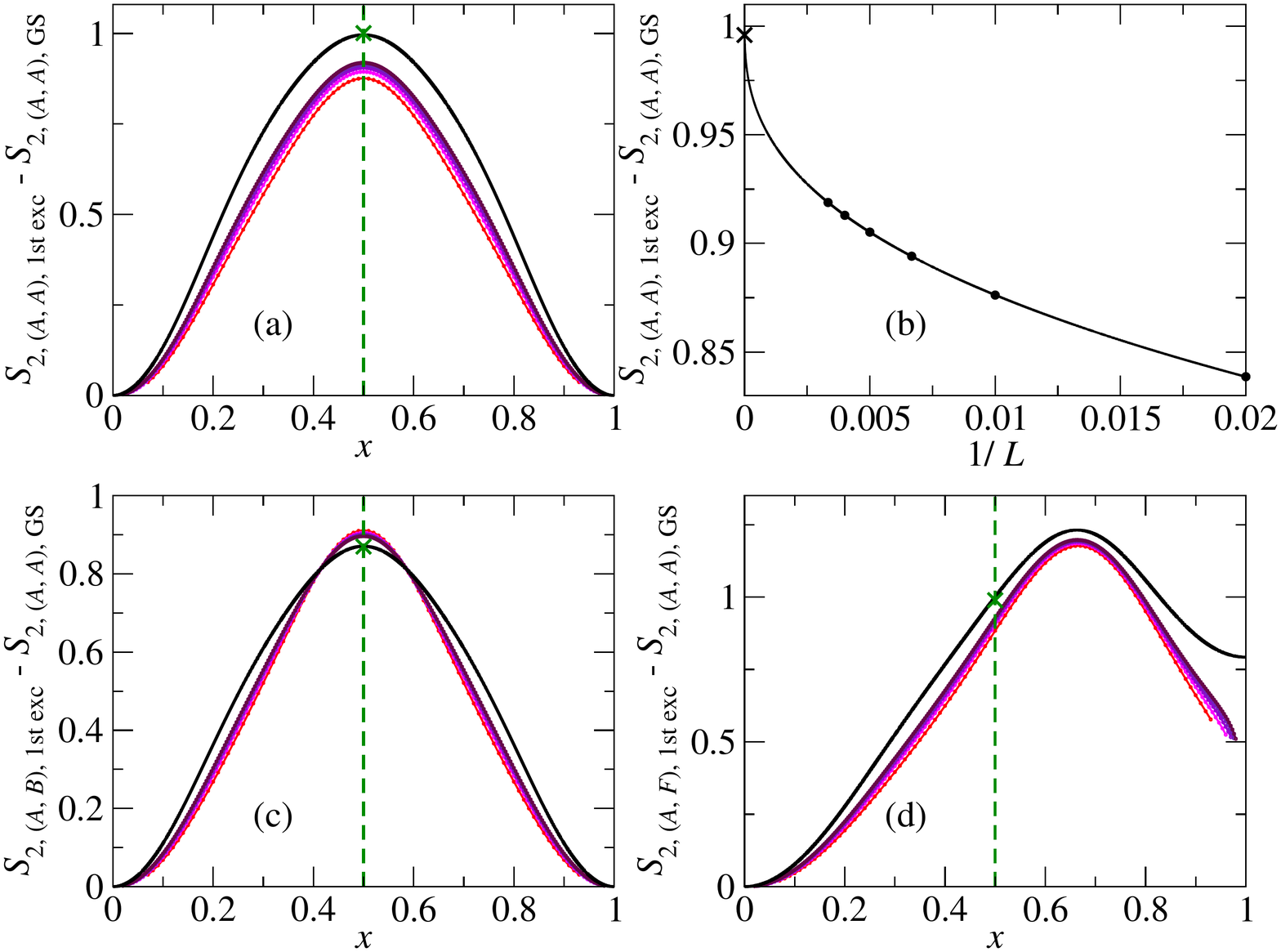}
\caption{$n=2$ EEE for descendant states of the critical three-state 
Potts chain with conformal OBC. Panels (a), (c)-(d): difference between $S_2$ 
for the considered descendant state and OBC (see Table 
\ref{table:PottsDescendants_fit} for details) and the GS for $(A,A)$BC, 
as a function of $x$ and for different values of $L$. Black line: CFT 
prediction; colored dotted lines: DMRG data (red to maroon: $L=100$ to $300$); 
vertical green dashed line: considered $x$($=1/2$) for the FSS analysis; green cross: 
result of the FSS analysis. Panel (b): FSS of the data of panel (a) at $x=1/2$. Circles: 
DMRG data; cross: CFT prediction; solid line: best fit, done by means of the 
three-parameters formula $y=a_0+a_1x^{a_2}$ (see Table 
\ref{table:PottsDescendants_fit} for the obtained values of the
coefficients and the deviations from the CFT 
predictions).}\label{PottsDescendants_fig}
\end{figure}

With the same choice of conformal OBC, we consider the first excited state, 
corresponding to the stress-energy tensor $T$. The CFT prediction for the 
$n=2$ EEE, obtained from Eq. (\ref{F_T}) by substituting $c=4/5$, looks
\begin{equation}
\fl F_T^{(2)}=\frac{578675+196776\cos(2\pi x)+36252\cos(4\pi x)+7448\cos(6\pi 
x)+49\cos(8\pi x)}{819200}.
\end{equation}
The comparison with the DMRG data is performed in Figure 
\ref{PottsDescendants_fig}(a): similarly to the case of the GS, the 
REE suffers from a strong finite-size correction, analogously to the $c=1/2$ behavior.
With a FSS approach similar to the one of Section \ref{Ising_sec}, we are able to 
show that the numerical data converges, in the thermodynamic limit, to the CFT 
prediction: this fact is shown in Figure \ref{PottsDescendants_fig}(b) (see Table 
\ref{table:PottsDescendants_fit} for quantitative details about the FSS 
procedure).
\begin{table}[t]
 \centering
 \begin{tabular}{|c|c|c|c|c|c|c|c|c|}
  \hline
  BC & $h$ & $a_0$ & $a_1$ & $a_2$ & $\eta$ & Figure\\
  \hline
  $(A,A)$ & $2$ & $1.001$ & $-0.720$ & $0.381$ & $5\times10^{-3}$ & 
\ref{PottsDescendants_fig}(a), (b)\\
  \hline
  $(A,B)$ & $5/3$ & $0.354$ & $-0.658$ & $0.415$ & $2\times10^{-3}$ & 
\ref{PottsDescendants_fig}(c)\\
  \hline
  $(A,F)$ & $9/8$ & $0.989$ & $-1.063$ & $0.502$ & $1.2\times10^{-2}$ & 
\ref{PottsDescendants_fig}(d)\\
  \hline
 \end{tabular}
 \caption{Fitting result of the finite-size data for the first descendant states 
of the three-state Potts chain with OBC. First column: considered boundary 
condition; second column: conformal weight of the considered state; third, 
fourth and fifth columns: estimated fit parameters (used formula: 
$y=a_0+a_1x^{a_2}$); sixth column: relative deviation $\eta$ of $a_0$ from the 
CFT prediction; seventh column: corresponding Figure. In any case, the analysis 
is performed at a subsystem size $x=1/2$.}\label{table:PottsDescendants_fit}
\end{table}
The agreement between the numerical data and the CFT prediction is 
remarkable.

We then consider the $(A,B)$BC case: the operator content of the theory consists in the tower of the primary field with $h=2/3$, that we dub $\psi$. In order to compute the $n=2$ EEE we need the four-point function of $\psi$ operators: we compute it relying on the parafermionic representation of the Potts model \cite{Zamolodchikov1985}. In this representation, the field $\psi$ is the parafermionic current itself, and its four-point function can be inferred from a recurrence relation. Here, importantly, $\psi$ is different from its adjoint. The final result reads
\begin{equation}\fl
\langle\psi(z_{1})\psi^\dagger(z_{2})\psi(z_{3})\psi^\dagger(z_{4})\rangle=z_{12
}^{-1/3}z_{13}^{-2/3}z_{14}^{-1/3}z_{32}^{-1/3}z_{24}^{-2/3}z_{34}^{-1/3}\left[
\frac{z_{14}z_{32}}{z_{12}z_{34}}+\frac{z_{12}z_{34}}{z_{14}z_{32}}-\frac{2}{3}
z_{24}\right].
\end{equation}
By means of this correlation 
function, the resulting EEE for the GS is easily computed:
\begin{equation}\label{psi0}
F^{(2)}_\psi=\frac{5+\cos(2\pi x)}{6}.
\end{equation}
For the first descendant state, applying the technique of  \ref{appB}, 
we obtain
\begin{equation}\label{psi1}
F^{(2)}_{L_{-1}\psi}=\frac{4558+1343\cos(2\pi x)+194\cos(4\pi x)+49\cos(6\pi 
x)}{6144}.
\end{equation}
Such predictions are compared with the DMRG data in Figs. 
\ref{PottsPrimaries_fig}(c) and \ref{PottsDescendants_fig}(c) (for the results 
of the FSS analysis, see Tables \ref{table:PottsPrimaries_fit} and 
\ref{table:PottsDescendants_fit}). Again, the agreement between the CFT 
prediction and the numerical data is remarkable. 
We point out that in the fusion rule $\psi\times\psi^\dagger=\mathbb{I}$ only the
identity channel is present: accordingly, a different
result for the same excited states but with different BC is not possible.

We conclude the Section with the $(A,F)$ case. In order to compute the $n=2$ 
EEE, we need the four-point function of the exotic primary field with 
scaling dimension $h=1/8$, that we call $\vartheta$. This field occupies the $(1,2)$ position in the Kac table and its four-point function can be calculated by the standard Coulomb-gas approach \cite{DiFrancesco1997}. For the considered OBC the result is
\begin{equation}
\eqalign{\langle\vartheta^{(A,F)}(z_{1})\vartheta^{(F,A)}(z_{2})\vartheta^{(A,F)}(z_{3}
)\vartheta^{(F,A)}(z_{4})\rangle= \cr
\qquad\qquad=z_{12}^{-1/4}z_{13}^{-5/12}z_{14}^{+5/12}z_{23}
^{+5/12}z_{24}^{-5/12}z_{34}^{-1/4}\,_{2}F_{1}\left(\frac{1}{6},\frac{5}{6}
;\frac{1}{3};\frac{z_{12}z_{34}}{z_{13}z_{24}}\right),}
\end{equation}
where $_2F_1(\alpha,\beta;\gamma;z)$ is the hypergeometric function. The conformal block above corresponds to the 
identity channel, that is the only one permitted by the fusion rule
\begin{equation}
\vartheta^{(A,F)}\times\vartheta^{(F,A)}=\mathbb{I}^{(A
,A)},
\end{equation}
which we read off from the partition function $Z_{A,A}$, Eq. (\ref{PottsAA}).
The EEE for the GS is then
\begin{equation}
F_\vartheta^{(2)}=\cos\left(\frac{\pi x}{2}\right)^{\frac{4}{3}} 
{_2F_1}\left(\frac{1}{6},\frac{5}{6};\frac{1}{3};y\right),
\end{equation}
and for the first descendant state we obtain, using Eq. (\ref{eq:L-1}),
\begin{equation}
\fl\eqalign{
F_{L_{-1}\vartheta}^{(2)}=\frac{1}{81}\cos\left(\frac{\pi 
x}{2}\right)^{\frac{4}{3}}\bigg[\left(101-156y+300y^2-568y^3+324y^4\right){_2F_1
}\left(\frac{1}{6},\frac{5}{6};\frac{1}{3}
;y\right)\cr
\qquad\qquad\qquad\qquad\qquad\qquad+\left(-20+44y+24y^2-80y^3+24y^4\right){_2F_1}\left(\frac{7}{6},\frac
{5}{6};\frac{1}{3};y\right)\bigg],}
\end{equation}
with $y=\sin(\pi x/2)^2$. We compare the analytical predictions and the DMRG 
data in Figs. \ref{PottsPrimaries_fig}(d) and \ref{PottsDescendants_fig}(d): 
after a FSS analysis (see Tables \ref{table:PottsPrimaries_fit} and 
\ref{table:PottsDescendants_fit} for quantitative details) we find, again, a nice
agreement between them.

Summarizing, we were able to find quantitative agreement between the CFT 
analytical predictions for the $n=2$ EEE's and the DMRG data for all of the 
considered conformal OBC.

\section{Compactified free boson and spin-$1/2$ XX chain}\label{XX}

The last CFT we consider is the free boson \cite{DiFrancesco1997}, described by 
the Lagrangian
\begin{equation}
\mathcal{L}=\frac{1}{8\pi}\int 
dx\left[(\partial_t\varphi)^2-(\partial_x\varphi)^2\right],
\end{equation}
compactified on a circle, i.e., $\varphi\simeq\varphi+2\pi R$,  $R$ being the 
compactification radius. This CFT is characterized by a central 
charge $c=1$ and is not minimal; the one-to-one correspondence between primary 
fields and conformal OBC is therefore not available. Indeed, it is possible to 
show that the conformal OBC are of two types, named Dirichlet ($D$) and Neumann 
($N$) \cite{Saleur1998}.

A simple lattice realization of the compactified free boson is the spin-1/2 XX 
chain in the vanishing-magnetization (half-filling) sector \cite{Giamarchi2003}. In order for it to 
realize the upper-half-plane CFT, the Hamiltonian to be considered is 
\cite{Bilstein2000}
\begin{equation}
H=-\frac{J}{2}\sum_j\left(\sigma_j^+\sigma_{j+1}^-+\sigma_j^-\sigma_{j+1}
^+\right)-b_1\sigma_1^x-b_L\sigma_L^x,
\end{equation}
where $\sigma_j^\pm=\left(\sigma_j^x\pm i\sigma_j^y\right)/2$; for $b_j/J=0$ 
($b_j/J\gg0$), $D$BC ($N$BC) are realized. There are three possible combinations of conformal BC: 
$DD$BC, $NN$BC and $ND$BC, respectively associated to the partition functions 
\cite{Saleur1998,Bilstein2000,Taddia2013}
\begin{eqnarray}
\eqalign{
Z_{DD}(q)&=\frac{1}{\eta(q)}\sum_{n\in\mathbb{Z}}q^{n^2/2},\\
Z_{NN}(q)&=\frac{1}{\eta(q)}\sum_{n\in\mathbb{Z}}q^{2n^2},\\
Z_{DN}(q)&=\frac{1}{2\eta(q)}\sum_{n\in\mathbb{Z}}q^{(n-1/2)^2/4}=\chi_{1/16}
(q)\left[\chi_0(q)+\chi_{1/2}(q)\right],}
\end{eqnarray}
where the last equality, proved in Ref. \cite{Taddia2013}, exploits the 
naive intuition of additivity of central charges \cite{DiFrancesco1997}. 
Once expanded around $q=0$, they look
\begin{eqnarray}
q^{c/24}Z_{DD}(q)&=1+2q^{1/2}+q+2q^{3/2}+4q^2+\mathcal{O}\left(q^{5/2}\right),
\label{Z_DD}\\
q^{c/24}Z_{NN}(q)&=1+q+4q^2+5q^{3}+\mathcal{O}\left(q^{4}\right),\label{Z_NN}\\
q^{c/24}Z_{DN}(q)&=q^{1/16}+q^{9/16}+q^{17/16}+2q^{25/16}+\mathcal{O}\left(q^{
33/16}\right)\label{Z_ND}.
\end{eqnarray}
In the theories described by Eqs. (\ref{Z_DD}) and (\ref{Z_NN}), the primary 
fields are the vertex operators $\normOrd{e^{\pm i \frac{n}{R}\varphi}}$, $n\in\mathbb{Z}$ at the compactification radii $R_{DD}=1$, $R_{NN}=1/2$ and the derivative of 
the field, $i\partial\varphi$. The contributions to the REE's originating from 
them have already been studied in Ref. \cite{Taddia2013}. In both cases, 
the first secondary operators have weight $2$, and come in multiplets; therefore,   
the problem of degeneracies cannot be avoided anymore. The case of Eq. (\ref{Z_ND}) 
has to be treated in a different way, i.e., by opportunely multiplying $c=1/2$ 
corrections, that have already been studied in Ref. \cite{Taddia2013} and 
in Section \ref{Ising_sec}.

The numerical data is produced by means of exact diagonalization for $DD$BC (in 
this case, the Hamiltonian can be reduced to a free spinless-fermions one by 
means of a Jordan-Wigner transformation \cite{Giamarchi2003}), and by means of the DMRG technique in 
the remaining cases. A chain of $L=200$ sites has been considered (the FSS 
analysis for the REE's is unnecessary in this case, for reasons that will be 
clear soon); 7 finite-size sweeps and up to $m=250$ Schmidt states have been 
employed; a maximum truncation error of $10^{-8}$ has been achieved in the last 
steps of the finite-system algorithm.

\begin{figure}[ht]
\centering
\includegraphics[width=\linewidth,trim=0.5cm 0.6cm 0cm 5.5cm, clip=true]{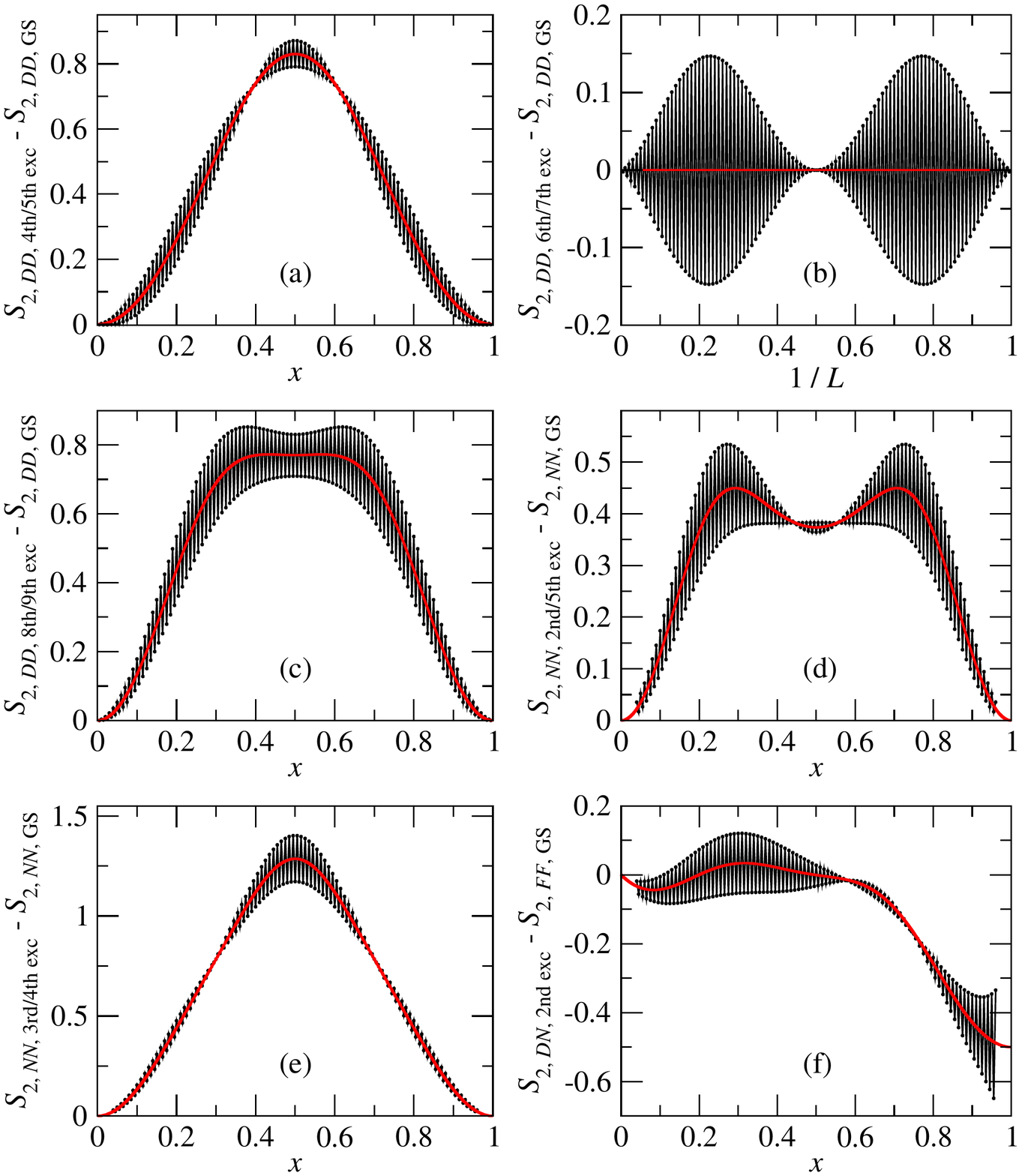}
\caption{$n=2$ EEE for descendant excited states of the spin-1/2 XX chain with 
conformal OBC. Black circles: numerical data; red lines: CFT 
predictions.}\label{XX_fig}
\end{figure}

We start by considering the case of $DD$BC: as shown by Eq. (\ref{Z_DD}), the 
excitations at levels $3/2$ and $2$ appear in multiplets. First, we consider the 
fourth and the fifth excited states, corresponding to the operators 
$L_{-1}\normOrd{e^{\pm i\varphi}}$. The method of  \ref{appB} allows to compute 
the corresponding $n=2$ EEE, starting from the well-known four-point 
correlations of vertex operators. The final result is
\begin{equation}
F^{(2)}_{L_{-1}\normOrd{e^{\pm i\varphi}}}=\frac{99+28\cos(2\pi 
x)+\cos(4\pi x)}{128},
\end{equation}
and is plotted, together with the numerical data, in Figure \ref{XX_fig}(a) 
(the numerically-computed REE is also the same for both states; however, for linear
combinations, it would be different).
The agreement between the two approaches is manifest, since the finite-size 
corrections oscillate as a function of $x$. This situation is typical of 
Luttinger liquids, i.e., free bosonic theories, as proven in many different situations (see, e.g., Refs. 
\cite{Laflorencie2006, Cardy2010, Calabrese2010, Dalmonte2011, Xavier2012, 
Dalmonte2012, Swingle2013,Cevolani2016}).

The sixth, seventh, eigth and nineth excitations are also degenerate, and possess the 
same conformal weight. However, the first and the second two display different 
behaviors, as shown in Figs. \ref{XX_fig}(b) and \ref{XX_fig}(c). In the CFT 
picture, there are four operators with conformal dimension $2$: $\normOrd{e^{\pm 
i2\varphi}}$, $T$ and $L_{-1}\partial\varphi$. In particular, the first two lead 
to vanishing EEE's \cite{Alcaraz2011,Berganza2012}. Comparing the 
prediction for them to the numerical data in Figure \ref{XX_fig}(b), we see that 
these two operators identify the sixth and the seventh excitation of the chain. 
Instead, the profiles of Figure \ref{XX_fig}(c) are well reproduced by the 
linear combinations of operators $T\pm L_{-1}\partial\varphi$, that lead to the 
$n=2$ EEE
\begin{equation}\fl
F^{(2)}_{T\pm L_{-1}\partial\varphi}=\frac{382659+106184\cos(2\pi 
x)+32924\cos(4\pi x)+2296\cos(6\pi x)+225\cos(8\pi x)}{524288}.
\end{equation}

In the case of $NN$BC, the degenerate descendant quadruplet is at level 2, and it is formed by
the operators $T$, $L_{-1}\partial\varphi$ and $\normOrd{e^{\pm i\varphi}}$. The linear combinations of operators that 
correctly interpret the DMRG data are $T\pm \normOrd{\cos\varphi}$ and 
$L_{-1}\partial\varphi\pm2i\normOrd{\sin\varphi}$.
Remarkably, the first linear combinations correspond to a bosonization of the 
stress-energy tensor of the $c=1/2$ free massless Majorana theory 
\cite{Kiritsis1987}: indeed, we were able to check exactly, by means of the general 
$n$-point function of vertex operators, that such combinations lead to the same
functional form, Eq. (\ref{F_T_Ising}), for the $n=2$ EEE as the stress-energy 
tensor in the $c=1/2$ minimal CFT. The CFT prediction is displayed, together with the DMRG 
data, in Figure \ref{XX_fig}(d) (the difference between the numerically-computed 
EEE's for the two states just resides in a relative minus sign in the 
coefficient of the oscillating part), showing remarkable agreement. Instead, 
the EEE for the other two states originates from a four-point function
involving both $i\partial\varphi$ and vertex operators. Once explicitly 
computed, it looks
\begin{equation}
\langle 
i\partial\varphi(z_1)i\partial\varphi(z_2)\normOrd{e^{i\varphi}}(z_3)\normOrd{e^{-i\varphi}
}(z_4)\rangle=\frac{1}{z_{12}^2z_{34}^4}+\frac{4}{z_{13}z_{14}z_{23}z_{24}z_{34}
^2},
\end{equation}
and the resulting EEE's are
\begin{equation}\fl
F^{(2)}_{L_{-1}\partial\varphi\pm2i\normOrd{\cos\varphi}}=\frac{\left[\cos (2 \pi  
x)+7\right] \left[1558+ 439 \cos (2 \pi  x)+26 \cos (4 \pi  x)+25 \cos (6 \pi  
x)\right]}{16384}.
\end{equation}
The successful comparison with the DMRG results is performed in Figure \ref{XX_fig}(e).

Finally, we consider $DN$BC. This case is different from the previous two, since 
the partition function can be written as a product of characters of the $c=1/2$ 
minimal CFT \cite{Taddia2013}. As shown by Eq. (\ref{Z_ND}), the first 
descendant state arises with a conformal weight $17/16$, and stems from the 
operator $\left(L_{-1}\sigma\right)\otimes\psi$. The $n=2$ EEE is 
therefore
\begin{equation}
F^{(2)}_{\left(L_{-1}\sigma\right)\otimes\psi}=F^{(2)}_{L_{-1}\partial\varphi}F^
{(2)}_\psi,
\end{equation}
as computed in Section \ref{Ising_sec} and in Ref. \cite{Taddia2013}. The 
comparison with the DMRG data is performed in Figure \ref{XX_fig}(f): the 
theoretical prediction and the numerical data match, up to the usual 
oscillating corrections and to an additive Affleck-Ludwig contribution, with 
$g=1/\sqrt{2}$, that is known to be associated to $N$BC \cite{Zhou2006}.

Concluding, even in this case we were able to show that the CFT low-energy 
picture captures the main features of the numerically computed EEE's.

\section{Conclusions and outlook}\label{conc}

In this work we extended the results of Ref. \cite{Palmai2014} on R\'enyi entanglement entropies of descendant states in conformal systems, to the case of conformal systems with boundaries. We provided a unifying approach applicable for both periodic and open boundary conditions and described the computation of the corrections to Eq. (\ref{CCREE}) for excited states. We also proved that for any choice of boundary conditions the R\'enyi entanglement entropies are formally given by the same $2n$-point functions; the only difference stems from the suppression of one of the chiralities and the different OPE coefficients realized with different boundary conditions (e.g., different conformal blocks solving the same differential equation will play a role with different boundary conditions).

Using this framework we explicitly computed the $n=2$ R\'enyi entanglement entropies for the first few excitation for three lattice models, belonging to different universality classes (Ising, XX or XXZ, and three-state Potts), and compared them with numerical data, obtained by means of the density-matrix-renormalization-group algorithm (and, where available, of free-fermions representations). In all the considered cases, the agreement between analytical predictions and numerical calculations was found to be excellent. Moreover, we were able to solve, for the first time, the problem of the R\'enyi entanglement entropies of degenerate energy multiplets, where different linear combinations are observed on the lattice and in the field theory.

The study of R\'enyi entanglement entropies in many-body systems offers several directions for future work. For instance, the understanding of finite-size corrections to the conformal scalings, that we have seen to arise in all of the analyzed situation, although with very different features, is still very incomplete, and deserves further investigation \cite{Laflorencie2006,Calabrese2010,Swingle2013}. In addition, the appearance of the Affleck-Ludwig constant contributions as part of the excess R\'enyi entanglement entropies is a phenomenon that has been observed since many years \cite{Zhou2006,Taddia2013}, and its complete theoretical understanding is still lacking. On the other side, R\'enyi entanglement entropies are among the most natural quantities that can be computed by means of the currently most popular numerical methods \cite{Schollwock2011,Humeniuk2012,Porter2016}. The present study could help in order to numerically identify the lattice realization of conformal boundary conditions: for the majority of critical lattice systems, the boundary conditions preserving conformal invariance are unknown. Moreover, our approach allows, in pronciple, for the numerical identification of the conformal fields corresponding to specific lattice states, especially for degenerate energy multiplets. We therefore think that our study can stimulate future activity in this fertile research field.

\section{Acknowledgements}

We thank G. Sierra for important discussions, for stimulating our interest in the project and for a careful reading of the manuscript; we thank M. I. Berganza and G. Tak\'acs for valuable discussions. We acknowledge INFN-CNAF for providing computational resources and support, and D. Cesini in particular; we thank S. Sinigardi for technical support. L. T. acknowledges financial support from the EU integrated project SIQS, while 
T. P. from the Hungarian Academy of Sciences through grant No. LP2012-50 and a postdoctoral fellowship.

\appendix

\section{Conformal transformation of generic fields}\label{appA}

We review in this Appendix the recipe, introduced in Ref. 
\cite{Gaberdiel1994}, needed in order to perform the conformal transformation of a 
generic field. For sake of clarity, only the chiral part of the 
field is considered.

Primary operators transform in a particularly simple way: the conformal mapping 
$w=f(z)$ takes the field $\Phi$, with conformal weight $h$, to 
\begin{equation}
U_{f}\Phi(z)U_{f}^{-1}=\left[f'(z)\right]^h\Phi(w).\label{eq:primary}
\end{equation}
The transformation rule for secondary fields is much less known, and much more complicated. It is given by
\begin{equation}
U_{f}\mathcal{O}(z)U_{f}^{-1}=\sum_{(p)}H_{(p)}[f,z)\left[L_{p_{1}}\ldots 
L_{p_{k}}\mathcal{O}\right](f(z)),
\end{equation}
where $L_{j}$ is the usual $j$-th Virasoro generator (relative to the origin), 
while $\mathcal{O}(z)$ is the operator corresponding to the state 
$\mathcal{O}(0)\vert0\rangle$, inserted at $z$. The coefficients $H_{(p)}[f,z)$ 
were identified, in Ref. \cite{Gaberdiel1994}, to be generated by the 
expression
\begin{equation}
\eqalign{
\prod_{n=0}^{\infty}e^{R_{n}[f,z)L_{n}}\mathcal{O}(0)\vert0\rangle&=\left[ 
\prod_{n=0}^{m}\sum_{k=0}^{\left\lfloor \frac{m}{n}\right\rfloor 
}\frac{\left(R_{n}L_{n}\right)^{k}}{k!}\right] 
\mathcal{O}(0)\vert0\rangle\cr&=\sum_{(p)}H_{(p)}[f,z)L_{p_{1}}\ldots 
L_{p_{k}}\mathcal{O}(0)\vert0\rangle,}
\end{equation}
where, in turn, the coefficients $R_{n}$ are defined recursively as 
\begin{eqnarray}
R_{0}(z) & = & \log f'(z),\\
R_{n}(z) & = & \frac{1}{n+1}(R'_{n-1}(z)-A_{n}(z)),\quad n\geq1,\nonumber 
\end{eqnarray}
with $A_{n}(z)$ given in Ref. \cite{Gaberdiel1994}, the first few being 
\begin{equation}
 A_{1}=0,\quad A_{2}=R_{1}^{2},\quad A_{3}=0,\quad 
A_{4}=\frac{3}{2}R_{2}^{2},\quad A_{5}=0,\quad A_{6}=R_2^3+2R_3^2.
\end{equation}
It is easy to check that $R_{2}$ is the Schwarzian derivative of $f$ multiplied 
by $1/3!$, reproducing the familiar transformation law of the stress-energy 
tensor (Eq. (\ref{transf_T})):
\begin{equation}
U_{f}T(z)U_{f}^{-1}=\left[f'(z)\right]^{2}T(f(z))+\frac{c}{12}\{f,z\}.
\end{equation}
Another example is the transformation law of the field $\partial\Phi$, with $\Phi$ a primary field of weigth $h$:
\begin{equation}\label{eq:L-1}
U_{f}\partial\Phi(z)U_{f}^{-1}=\left[f'(z)\right]^{h+1}
\partial\Phi(f(z))+hf''(z)\left[f'(z)\right]^{h-1}\Phi(f(z)).
\end{equation}
The relation above can independently be deduced from Eq. (\ref{eq:primary}) using the chain 
rule of differentiation.

\section{Evaluation of $N$-point functions of descendant fields}\label{appB}

We review in this Appendix the strategy, developed in Ref. \cite{Palmai2014}, for evaluating a generic $N$-point correlation function of
chiral secondary fields.

The basic object that is considered is
\begin{equation}\label{basicobject}
\left\langle (L_{n}A_{1})(z_{1})A_{2}(z_{2})\ldots A_{N}(z_{N})\right\rangle,
\end{equation}
where $A_j$ is a generic secondary field, generated from a primary by the
application of Virasoro generators. Ultimately, as shown below, such $N$-point function can be transformed into a sum of derivatives of the corresponding $N$-point function of primary fields. 

In the first part of the procedure the generator $L_n$ is removed from the $A_1$ operator alone. In order to perform this task, the contour-integral representation of the Virasoro generators is used \cite{DiFrancesco1997}:
\begin{equation}
L_{n}\mathcal{O}(z)=\oint_{z}\frac{d\zeta}{2\pi 
i}(\zeta-z)^{n+1}T(\zeta)\mathcal{O}(z),
\end{equation}
where the subscript $z$ indicates that the contour encircles $z$. After 
inserting this expression in Eq. (\ref{basicobject}), the contour 
of integration can be deformed in order to enclose all the other poles of this integral, i.e., the $z_j$ 
insertion points. This gives
\begin{equation}
\eqalign{
 &\left\langle \oint_{z_{1}}\frac{d\zeta}{2\pi 
i}(\zeta-z_{1})^{n+1}T(\zeta)A_{1}(z_{1})A_{2}(z_{2})\dots 
A_{N}(z_{N})\right\rangle=\cr
& \qquad=-\left\langle A_{1}(z_{1})\left[\oint_{z_{2}}\frac{d\zeta}{2\pi 
i}(\zeta-z_{1})^{n+1}T(\zeta)A_{2}(z_{2})\right]\dots A_{N}(z_{N})\right\rangle- 
\cr
& \qquad\phantom{=} -\left\langle 
A_{1}(z_{1})A_{2}(z_{2})\left[\oint_{z_{3}}\frac{d\zeta}{2\pi 
i}(\zeta-z_{1})^{n+1}T(\zeta)A_{3}(z_{3})\right]\dots A_{N}(z_{N})\right\rangle-
 \ldots
}
\end{equation}
Now, using the relation
\begin{equation}
(\zeta-z_{1})^{n+1}=\sum_{k=0}^{\infty}{{n+1}\choose{k}}(z_{i}-z_{1})^{n+1-k}(\zeta-z_{i})^{k},
\end{equation}
and so on, the integrals can be removed and replaced by Virasoro generators acting on the operators $A_{j>1}$:
\begin{equation}\fl
\eqalign{
&\left\langle (L_{n}A_{1})(z_{1})A_{2}(z_{2})\ldots A_{N}(z_{N})\right\rangle =
\cr
&\qquad=-\sum_{k=-1}^{\infty}{{n+1}\choose{k+1}}(z_{2}-z_{1})^{n-k}\left\langle 
A_1(z_{1})(L_{k}A_{2})(z_{2})\dots A_{N}(z_{N})\right\rangle-\cr
&\qquad\phantom{=}-\sum_{k=-1}^{\infty}{{n+1}\choose{k+1}}(z_{3}-z_{1})^{n-k}
\left\langle A_1(z_{1})A_2(z_{2})(L_kA_3)(z_{3})\dots 
A_{N}(z_{N})\right\rangle-\ldots
}
\end{equation}
Iterating this procedure, all the Virasoro generators can be removed from $A_1$, 
paying the price of complicating the remaining operators in the correlator: indeed, 
what is obtained is a finite sum of correlators, with the operator inserted at 
$z_1$ being primary.

The next step consists in repeating the procedure above for $A_2$, reducing it to a primary. However, at this point, Virasoro generators with $n=-1,\;0$ will appear again in front of the first operator, potentially followed by other Virasoro generator of any order: this apparently could make all the previous efforts useless. Actually, this is not the case, because of the following identity: 
\[
L_{n}L_{-1}^{m}\Phi=c_{nm}L_{-1}^{m-n}\Phi,\qquad n\geq-1;
\]
$\Phi$ is a primary field, indicating that the addition of Virasoro generators in front of powers of 
$L_{-1}$ results in a sequence of $L_{-1}$ generators, i.e., a partial derivative. 
The final result is then
\begin{equation}
\sum_{j}C_{j}(z_{1},\ldots,z_{n})\partial_{z_{1}}^{m_{1j}}\partial_{z_{2}}^{m_{
2j}}\ldots\partial_{z_{N-1}}^{m_{N-1,j}}\left\langle 
\Phi_{1}(z_{1})\Phi_{2}(z_{2})\ldots\Phi_{N}(z_{N})\right\rangle,
\end{equation}
with proper $C_j(z_1,\ldots,z_N)$ coefficients and $\Phi_j$ all primaries. Therefore, everything is reduced 
to the computation of sums of derivatives of $N$-point functions of primaries, 
computable by means of, e.g., the Coulomb gas formalism \cite{DiFrancesco1997} in the case of minimal models.

\end{document}